\documentclass[aps,pra,twocolumn,superscriptaddress,showpacs,floatfix]{revtex4-1}
\usepackage{graphicx,amsmath,amssymb,amsfonts,xspace,epsfig,float,multirow}
\usepackage{amsfonts}
\usepackage{braket}
\usepackage{epstopdf,xcolor}
\usepackage{bm}
\usepackage[caption=false]{subfig}

\usepackage[english]{babel}

\usepackage{hyperref}

\newcommand*\dd{\mathop{}\!\mathrm{d}}

\newcommand{\eM}{\mathcal{M}}
\newcommand{\eN}{\mathcal{N}}

\begin{document}
	
	\title{Multi-mode Bose-Hubbard model for quantum dipolar gases in confined geometries}
	\author{Florian Cartarius}
\affiliation{Universit\'e Grenoble-Alpes, CNRS, Laboratoire de Physique et Mod\'elisation des Milieux Condens\'es, 38000 Grenoble, France}
	\affiliation{Theoretische Physik, Universit\"at des Saarlandes, D66123 Saarbr\"ucken, Germany}
	\author{Anna Minguzzi}
\affiliation{Universit\'e Grenoble-Alpes, CNRS, Laboratoire de Physique et Mod\'elisation des Milieux Condens\'es, 38000 Grenoble, France}
	\author{Giovanna Morigi}
	\affiliation{Theoretische Physik, Universit\"at des Saarlandes, D66123 
		Saarbr\"ucken, Germany}
	
	\begin{abstract}
We theoretically consider ultracold polar molecules in a wave guide. The particles are bosons, they experience a periodic potential due to an optical lattice oriented along the wave guide and are polarised by an electric field orthogonal to the guide axis. The array is mechanically unstable by opening the transverse confinement in the direction orthogonal to the polarizing electric field and can undergo a transition to a double-chain (zigzag) structure. For this geometry we derive a multi-mode generalized Bose-Hubbard model for determining the quantum phases of the gas at the mechanical instability taking into account the quantum fluctuations in all directions of space. Our model limits the dimension of the numerically relevant Hilbert subspace by means of an appropriate decomposition of the field operator, which is obtained from a field theoretical model of the linear-zigzag instability. We determine the phase diagrams of small systems using exact diagonalization and find that, even for tight transverse confinement, the aspect ratio between the two transverse trap frequencies controls not only the classical but also the quantum properties of the ground state in a non-trivial way. Convergence tests at the linear-zigzag instability demonstrate that our multi-mode generalized Bose-Hubbard model can catch the essential features of the quantum phases of dipolar gases in confined geometries with a limited computational effort. 
	\end{abstract}
	\date{\today}

	\maketitle

	\section{Introduction}

Dipolar bosonic gases offer a laboratory for studying the interplay of finite-range interactions and quantum fluctuations \cite{Lahaye,Jin}. The study of their dynamics in optical lattices, moreover, allows one to realize and characterize strongly-correlated states of ultracold matter \cite{Lahaye,Jin,BlochRMP,Ferlaino2016}. The essential features of the quantum phases of ultracold dipoles in optical lattices are believed to be captured by the so-called extended Bose Hubbard Model \cite{EBH}. This model reduces to the single-band Bose-Hubbard model for vanishing dipolar coupling, which at commensurate densities exhibits the Mott-Insulator to Superfluid quantum phase transition \cite{Bose-Hubbard,BlochRMP,FisherAndFisher}. For finite strengths of the dipolar interactions, in addition, it includes a finite-range interaction term that favours the appearance of diagonal long-range order \cite{Goral,Menotti,Batrouni2013,Deng2013a,Deng2013b,Batrouni2014}. 

\begin{figure}
\centering
\includegraphics[width=0.55\textwidth]{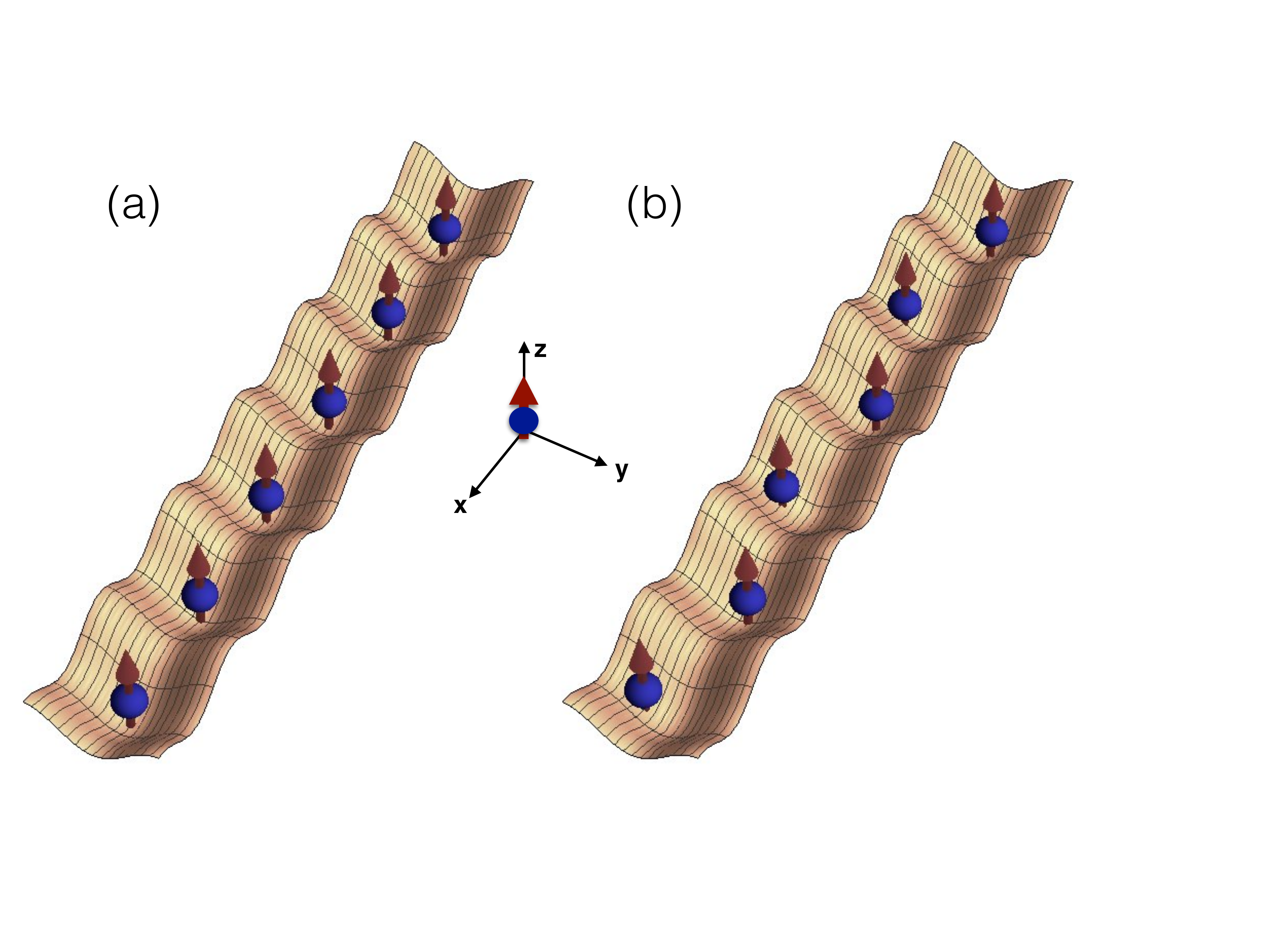}
\caption{(color online)  (a) Ultracold dipolar gases in an optical lattice along $x$ form an array when the confinement in the $y-z$ plane is sufficiently tight. (b) The dipoles form a zigzag chain when the trap frequency along $y$ is below a critical value and the dipoles are aligned along the $z$ axis orthogonal to the plane where the transition occurs. Starting from the array in the $x$ direction we develop a multi-mode Bose-Hubbard model which describes the onset of this classical structure, thus treating the transverse displacement as a continuous variable, while systematically accounting for quantum fluctuations along all directions in spaces. \label{Fig:1}}
\end{figure}

In three dimensions the anisotropic nature of the dipolar interaction is reflected in the properties of the Bose-Hubbard coefficients and can be analysed by orienting the dipolar structure by means of an external field \cite{Ferlaino2016}. When the motion is confined on a plane, instead, the mutual dipolar interaction can be made effectively isotropic and repulsive by orienting the dipoles perpendicularly to the plane itself. In this regime crystalline structures can emerge from the competition between the external confinement and the particles repulsion \cite{Astrakharchik:2007,Buechler:2007,Jin}. 

One exemplary situation is the linear-zigzag instability. This instability can be observed by tuning the frequency of the transverse trap, confining the dipoles along the array, and is illustrated in Fig. \ref{Fig:1} for a chain of dipoles in an optical lattice. For an incompressible chain the transition is continuous and the classical order parameter is the transverse displacement \cite{Fishman,Astrakharchik,Altman}, while the quantum linear-zigzag transition is of the same universality class as the Ising model in transverse field \cite{Shimshoni}. When the chain is compressible, instead, the classical transition becomes of weak first order \cite{Cartarius}, while the corresponding quantum behaviour is yet unexplored. In these respects the model we consider is peculiar, since the compressibility results from the interplay between interactions and quantum fluctuations and can be thus tuned by changing the lattice depth of the transverse confinement. Furthermore, previous literature pointed out that quantum fluctuations in the transverse directions can substantially modify the effective interaction the dipoles experience along the axis \cite{Sinha,Deuretzbacher,Recati,Sowinski}. The description of the structural instability, therefore, requires the development of a suitable model which describes spatial selforganization in the transverse direction, while the dipoles density is periodically modulated along $x$ and quantum fluctuations in all directions of space are appropriately taken into account. 

In this work we systematically derive a multi-mode extended Bose-Hubbard (EBH) model which is particularly apt to describe the phase diagram deep in the linear chain as well as close to the linear-zigzag instability. Our model is derived by identifying a suitable basis for the transverse excitations, which is obtained using the field theoretical description of the linear-zigzag instability \cite{Shimshoni,Silvi,Podolsky}. The dynamics takes into account the anisotropic nature of the dipolar interaction by calculating the integrals defining the EBH coefficients in three dimensions, thus including the fluctuations in the three-dimensional space. 

This manuscript is organized as follows. In Section \ref{Sec:1} we report the detailed derivation of the multi-mode extended Bose-Hubbard model. In Section \ref{Sec:2} we determine the phase diagrams of small systems for different aspect ratios of the transverse confinement using exact diagonalization. Moreover, we test the convergence of our basis choice at the linear-zigzag instability. The conclusions are drawn in Sec. \ref{Sec:Conclusions} while the Appendix reports calculations complementing the material of Sec. \ref{Sec:1}.

\section{Derivation of the multi-mode Bose-Hubbard model}
\label{Sec:1}
	
We consider a gas of identical dipolar molecules with mass $M$ and dipolar moment ${\bm p}$, interacting via the dipolar potential $U_d(\bm r) $ (with $\bm r=(x,y,z)$ the distance between the centers of mass of two molecules):
\begin{equation}
	U_d(\bm r) = \frac{p^2}{r^3} - \frac{3(\bm p \cdot \bm r)^2}{r^5} \,.
\end{equation}
An external electric field, moreover, aligns the dipoles along the $z$ direction. The molecules form an array along the $x$-axis due to the tight confinement of an external harmonic trap, 
\begin{equation}
V_{\rm trap}(y,z) = \frac 1 2 M (\omega_y^2 y^2 + \omega_z^2 z^2)\,,
\end{equation}
where the trap frequencies $\omega_y$ and $\omega_z$ are chosen such that $\omega_y\ll\omega_z$. The frequency $\omega_y$ is assumed to take value close to the critical value $\omega_y^{(c)}$, at which the linear-zigzag instability occurs in the mean-field model \cite{Astrakharchik, Shimshoni}. 

The dipoles are ultracold and obey the Bose-Einstein statistics. They also interact via $s$-wave van-der-Waals collisions and occupy the lowest bands of an optical lattice along the $x$ direction, 
\begin{equation}
V_{\rm opt}(x)=V_L \cos^2(\pi x/a)\,,
\end{equation}
where $V_L$ is the lattice depth and $a$ the lattice constant. Their state is described in second quantization by means of the bosonic field operators $\Psi(\bm r)$, $\Psi(\bm r')^\dagger$, with $[\Psi(\bm r),\Psi(\bm r')^\dagger]=\delta^{(3)}(\bm r-\bm r')$ and is governed by Hamiltonian $\mathcal H$, which reads 
	\begin{align}
	\mathcal H =& \int \dd{}^3r\, \Psi^\dagger(\bm r)  \left[ -\frac{\hbar^2}{2M} \nabla^2 + V_{\rm tot} (\bm r) \right]  \Psi(\bm r) \nonumber \\
	&+ \frac 1 2 \int \dd{}^3r \int \dd{}^3r' \,  \Psi^\dagger(\bm r) \Psi^\dagger(\bm r') U(\bm r - \bm r')   \Psi(\bm r') \Psi(\bm r) \,,
	\label{eq:hamilton}
	\end{align}
where $ V_{\rm tot} (\bm r)=V_{\rm trap}(y,z) +V_{\rm opt}(x)$. The interaction potential is the sum of the dipolar and of the contact interaction: $$ U(\bm r) =  U_d(\bm r) + U_g(\bm r)\,,$$
where $U_g(\bm r) = g \delta^{(3)}(\bm r)$ describes the $s$-wave scattering contribution, with $g=4 \pi \hbar^2 a_S/ M$ and $a_S$ the $s$-wave scattering length.	
			
\subsection{Mode expansion of the bosonic field operator}

In order to derive a convenient multi-mode EBH model we use a suitably-chosen mode expansion. We first assume that the molecules are tightly bound at the minima of the optical lattice and we perform the single-band approximation. We thus denote by $w_j(x)$ the real-valued Wannier function at site $j$ for the motion of a particle of mass $M$ moving along $x$ and experiencing the potential $V_{\rm opt}(x)$. The motion along the $z$ axis is assumed to be in the ground state of the harmonic oscillator at frequency $\omega_z$ with wave function $\theta_{0}(z)$:
\begin{equation}
\theta_0(z)=\frac{1}{\sqrt{\sqrt{\pi}\sigma_z}}\exp\left(-\frac{z^2}{2\sigma_z^2}\right)\,,
\end{equation}
and $\sigma_z=\sqrt{\hbar/(M\omega_z)}$.  
The motion along $y$ is instead decomposed into the basis $\{\phi_{m}(y)\}$ which diagonalizes an effective local Hamiltonian along $y$ according to a procedure first developed in Ref. \cite{Silvi}. This effective Hamiltonian includes the harmonic oscillator in the $y$ direction as well as the effective potential along $y$ due to the dipolar interactions. At the linear-zigzag structural transition the Hamiltonian describes an effective $\varphi^4$ model on a lattice, where the transition point at fixed linear density is given by the transverse trap frequency $\omega_y^{(c)}$. In detail, at site $j$ the Hamiltonian reads
\begin{equation}
\label{H_loc}
H_{\text{loc}}^{(j)}=- \frac{\hbar^2}{2M} \frac{\partial^2}{\partial y_j^2} +\frac{1}{2}M\omega_y^2y_j^2 +U_{\rm pin}^{(j)}\,,
\end{equation}
where $U_{\rm pin}^{(j)}$ is the local component of the dipole-dipole interaction $U_{\rm pin}$ describing the motion along $y$ assuming the dipoles are pinned at the classical equilibrium positions in the $x-z$ plane. Specifically,
\begin{align*}
U_{\rm pin}  =& \frac{p^2}{2} \sum_{[j \neq l]} \frac{1}{\left((j-l)^2a^2 +(y_j-y_l)^2 \right)^{3/2}} \,,
\end{align*}
and $U_{\rm pin}=\sum_jU_{\text{pin}}^{(j)}+\sum_{[j \neq l]} H_{\text{int}}^{(j,l)}$. The effective $\varphi^4$ model is found close to the structural instability, where $|y_j|\ll1$, when discarding the coupling with the axial modes due to the term $U_{\rm comp}=U_d -U_{\rm pin}$. In this limit the potential of Eq. \eqref{H_loc} can be cast in the form \cite{Shimshoni,Silvi}:
\begin{equation}
\label{eq:localhamintext}
\frac{1}{2}M\omega_y^2y_j^2 +U_{\rm pin}^{(j)}\simeq   \frac{1}{2}\left(M\omega_y^2 - \frac{p^2} {a^5}\eM_1\right) y_j^2 +
	\frac{p^2} {2a^7}\eM_2\; y_j^4\,,
\end{equation}
while the relevant terms of the sum $\sum_{[j \neq l]} H_{\text{int}}^{(j,l)}$ are $H_{\text{int}}^{(j,j+1)}=p^2/(2a^5)\eN_1 \left( y_j + y_{j+1} \right)^2$ and describe an effective nearest-neighbour interaction. For completeness, we report the explicit form of the dimensionless coefficients: $\eN_{1} =(9/4)\zeta(3)$, and 
$$\eM_{q=1,2}=\frac{\left(2^{3+2q} - 1 \right) \Gamma(q + \textstyle{\frac{3}{2}})}{
 q! \;4 \; \Gamma(\textstyle{\frac{3}{2}})}\, \zeta(3+2q)\,,$$
with $\zeta(\ell)$ Riemann's zeta function and $\Gamma(z)$ the Gamma's function \cite{Abramowitz}. From potential \eqref{eq:localhamintext} one directly determines the mean-field critical frequency $\omega_y^{(c)}$, at which the chain becomes mechanically unstable \cite{Astrakharchik,Shimshoni,Silvi}:
\begin{equation}
\omega_y^{(c)}=\sqrt{\eM_1 p^2/(Ma^5)}\,.
\end{equation}
	
We note that the basis $\{\phi_{m}(y)\}$ is found by numerically diagonalizing Hamiltonian $H_{\text{loc}}^{(j)}$ at site $j$, without performing any Taylor truncation of the local potential $U_{\rm pin}^{(j)}$. For $\omega_y\gg\omega_y^{(c)}$ we checked that it is well approximated by the eigenbasis of the harmonic oscillator at frequency $\omega_y$. For $\omega_y\simeq \omega_y^{(c)}$, instead the eigenbasis significantly differs from the oscillator eigenstates \cite{Silvi}. 

Using these prescriptions we decompose the field operator as 
\begin{align}
\label{eq:wfun}
\Psi(\bm r) = \sum_{j,m,n} w_j(x) \phi_m(y) \theta_0(z) a_{j,m},
\end{align}
where  $a_{j,m}$ is the bosonic operator which annihilates a particle at site $j$ and in the local quantum state $|m\rangle$, and
$[a_{j,m},a^{\dagger}_{\ell,n}]=\delta_{j,\ell}\delta_{n,m}$. 

\subsection{Multi-mode Bose-Hubbard model}

The multi-mode EBH model $H_{BH}$ for our study is obtained by substituting Eq. \eqref{eq:wfun} in the field operators of Hamiltonian \eqref{eq:hamilton}, by integrating out the position variables and by keeping only nearest-neighbor interactions. The resulting EBH model  exhibits a number of terms of different origin, which are conveniently identified by writing $H_{BH}$ as 
\begin{eqnarray}
	\label{H:BH}
	H_{BH} = H^x + H^y + H^{xy}\,,
\end{eqnarray}
where the three terms describe the axial and transverse motion, as well as their mutual interaction, respectively.

\begin{widetext}
\subsubsection{The axial motion}

The axial motion can be cast into the sum over the transverse bands labeled by the quantum number $m$, $H^x=\sum_mH^x_m$, with 
\begin{eqnarray}
	H^x_m&=& \epsilon^x\sum_j n_{j,m}-J^x\sum_{j}\left( a_{j,m}^\dagger a_{j+1,m}+{\rm H.c.}\right)+\frac{U^x_{m}}{2}\sum_{j}n_{j,m}(n_{j,m}-1)+V^x_{m}\sum_{j}n_{j,m}n_{j+1,m}\nonumber\\
	&  &+\frac{P^x_m}{2}\sum_{j}\left( a_{j,m}^\dagger a_{j,m}^\dagger a_{j+1,m}a_{j+1,m}+{\rm H.c.}\right)-\frac{T^x_m}{2}\sum_{j}\left(a_{j,m}^\dagger (n_{j,m} +n_{j+1,m})a_{j+1,m}+{\rm H.c.}\right)\,,
\label{H:x}
\end{eqnarray}
where we used a notation which can be put in direct connection with the EBH model of Ref. \cite{Sowinski} for the single band case ($m=0$). Here, $n_{j,m}=a_{j,m}^\dagger a_{j,m}$ denotes the particle number at site $j$ and with quantum number $m$. The first three terms on the right-hand side (RHS) are the onsite energy, with $\epsilon^x$ the single-particle energy in the lattice, the hopping along the axis scaled by the hopping coefficient $J_x$, and the onsite interaction including the contribution of the dipolar term. Their explicit form is 
\begin{eqnarray}			
			&&\epsilon^x=\int dx \, w_j(x) \left[-\frac{\hbar^2}{2 M} \frac{\partial^2}{\partial x^2} + V_L \cos^2 \left(\frac{\pi x}{a}\right) \right]w_{j}(x)\,, \\
			&&J^x=\int dx \, w_j(x) \left[\frac{\hbar^2}{2 M} \frac{\partial^2}{\partial x^2} - V_L \cos^2 \left(\frac{\pi x}{a}\right) \right]w_{j+1}(x)\,,\\
\label{BH:coeff:x1}
			&&U^x_{m}= \int d^3r_1 \int d^3r_2 U(\bm r_1 - \bm r_2) w_j^2(x_1)w_j^2(x_2)  \phi^{*2}_m(y_1)\phi^2_m(y_2) \theta^2_0(z_1) \theta^2_0(z_2) \,.
\end{eqnarray}
All other terms are solely due to dipole-dipole interaction and are the dipole blockade, whose strength is scaled by the coefficient $V^x_{m}$, the pair-hopping term, scaling with $P^x_m$, and the density-dependent tunnelling, proportional to $T^x_m$. These latter coefficients depend on the transverse quantum state $m$ and read
\begin{eqnarray}
\label{BH:coeff:x2}
			&& V_m^x = \int d^3r_1 \int d^3r_2 \,U_d(\bm r_1 - \bm r_2) w_j^2(x_1)w_{j+1}^2(x_2)|\phi_m(y_1)\phi_m(y_2)|^2\theta_0^2(z_1)\theta_0^2(z_2)\,,\\
\label{BH:coeff:x3}
			&&P^x_m =  \int d^3r_1 \int d^3r_2 \,U_d(\bm r_1 - \bm r_2) w_j(x_1) w_{j+1}(x_1) w_j(x_2) w_{j+1}(x_2) |\phi_m(y_1)\phi_m(y_2)|^2\theta_0^2(z_1)\theta_0^2(z_2)\,, \\
\label{BH:coeff:x4}
			&& T^x_m =-\int d^3r_1 \int d^3r_2 \,U_d(\bm r_1 - \bm r_2) w_j^2(x_1)w_{j}(x_2)w_{j+1}(x_2)|\phi_m(y_1)\phi_m(y_2)|^2\theta_0^2(z_1)\theta_0^2(z_2)\,.
\end{eqnarray}
When the transverse motion is in the ground state, namely, for $m=0$, Hamiltonian $H^x_m$ reduces to the model studied in Ref. \cite{Sowinski}. If in addition one discards the pair-hopping and the density-dependent tunneling terms, then $H^x_m$ corresponds to the so-called extended Bose-Hubbard model, whose phase diagram has been extensively analysed in Refs. \cite{Berg,Batrouni2013,Batrouni2014}.

\subsubsection{Transverse motion} 

The EBH term for the Hamiltonian governing solely the motion along $y$ takes the form $H^y=\sum_jH^y_j$ and is local in the site $j$. Each term of the sum reads
		\begin{eqnarray}
		H^y_j&=& \sum_m \epsilon_{m}^y n_{j,m} - \sum_{m\neq n} J^y_{m,n} a_{j,m}^\dagger a_{j,n} +\frac{1}{2}{\sum_{l,m,n,q}}^\prime U^y_{l,m,p,q}a_{j,l}^\dagger a_{j,m}^\dagger a_{j,n}a_{j,q} \, ,
		\label{H:y}
		\end{eqnarray}
where ${\sum_{l,m,n,q}}^\prime$ indicates that at least one of the indices $l,m,p,q$ is different from the others. The coefficients are independent of the lattice site $j$ since the Hamiltonian is invariant per discrete translation (for periodic boundary conditions). Here, the eigenenergy $\epsilon_{m}^y$ and the tunneling term $J^y_{m,n}$ read
\begin{eqnarray}			
			&&\epsilon^y_m=\int dy \, \phi_m(y)^*  \left[-\frac{\hbar^2}{2 M} \frac{\partial^2}{\partial y^2} + V_{\rm trap}(y) \right]\phi_m(y)\,, \\
			&&J^y_{mn}=\int dy \, \phi_m(y)^*  \left[\frac{\hbar^2}{2 M} \frac{\partial^2}{\partial y^2} - V_{\rm trap}(y) \right]\phi_n(y)\,,
\end{eqnarray}
while the interaction term $U^y_{l,m,p,q}$ takes the form
\begin{eqnarray}
\label{BH:coeff:y}
			U^y_{l,m,p,q}= \int d^3r_1 \int d^3r_2 U(\bm r_1 - \bm r_2) w_j^2(x_1)\theta_0^2(z_1) w_j^2(x_2)\theta_0^2(z_2) \phi_l^*(y_1)\phi_m(y_2)\phi_n^*(y_1)\phi_q(y_2)\,.
\end{eqnarray}
We remark that also the onsite term $U^y_{m,m,m,m}= U^x_m$ contributes in determining the transverse motion. We arbitrarily assigned this term to the axial EBH Hamiltonian $H^x$ and did not include it in Eq. \eqref{H:y} in order to avoid double-counting in the resulting EBH Hamiltonian $H_{BH}$, Eq. \eqref{H:BH}. 

\subsubsection{Coupling between axial and transverse degrees of freedom} 

Finally, $H^{xy}$ describes the interaction between excitations along the $x$ and the $y$ direction, it is solely due to the dipolar interaction and can be written as 
\begin{eqnarray}
\label{H:xy}
H^{xy}=\sum_{j,m}H^{xy}_{j,m}
\end{eqnarray}
where
\begin{eqnarray}
	H^{xy}_{j,m}=\frac{1}{2}{\sum_{\ell_1,\ell_2,\ell_3}}^{\prime}{\sum_{n_1,n_2,n_3}}^{\prime} V^{j,\ell_1,\ell_2,\ell_3}_ {m,n_1,n_2,n_3} a_{j,m}^\dagger a_{\ell_1,n_1}^\dagger a_{\ell_2,n_2}a_{\ell_3,n_3} \,,	
	\label{H:cross}
\end{eqnarray}
and describes a four-vertex type of interaction. We note that ${\sum_{\ell_1,\ell_2,\ell_3}}^{\prime}$ (${\sum_{n_1,n_2,n_3}}^{\prime}$) means that at least one of the indices $\ell_1,\ell_2,\ell_3$ has to be different from $j$ (respectively, at least one of the indices $n_1,n_2,n_3$ has to be different from $m$). Due to the tight-binding assumption, $\ell_1,\ell_2,\ell_3=j,j+1$ or $\ell_1,\ell_2,\ell_3=j,j-1$. The coefficients are found by performing the integral:
\begin{eqnarray}
\label{BH:coeff:xy}
V^{j,\ell_1,\ell_2,\ell_3}_{m,n,q,r}= \int d^3r_1 \int d^3r_2 \,U_d(\bm r_1 - \bm r_2) w_j(x_1)w_{\ell_1}(x_1)w_{\ell_2}(x_2)w_{\ell_3}(x_2)\phi_m^*(y_1)\phi_{n_1}^*(y_2)\phi_{n_2}(y_2)\phi_{n_3}(y_1)\theta_0^2(z_1)\theta_0^2(z_2)	\,.\end{eqnarray} 
\end{widetext}
Term $H^{xy}$ contains two physically relevant contributions. One contribution leads to the interaction term $H_{\text{int}}^{(j,j+1)}$ of the $\varphi^4$ model. The other is a coupling between axial and transverse modes, which becomes relevant when the chain is compressible \cite{Cartarius}. When the chain is incompressible, for hard-core bosons and unit filling the action of Hamiltonian $H^y+H^{xy}$ can be reduced to an effective  $\varphi^4$ model and the linear-zigzag transition is of the same universality class of the Ising model in transverse field \cite{Shimshoni,Silvi}.

\begin{figure*}
		\centering
		\subfloat[]{
			\includegraphics[width=0.32\textwidth]{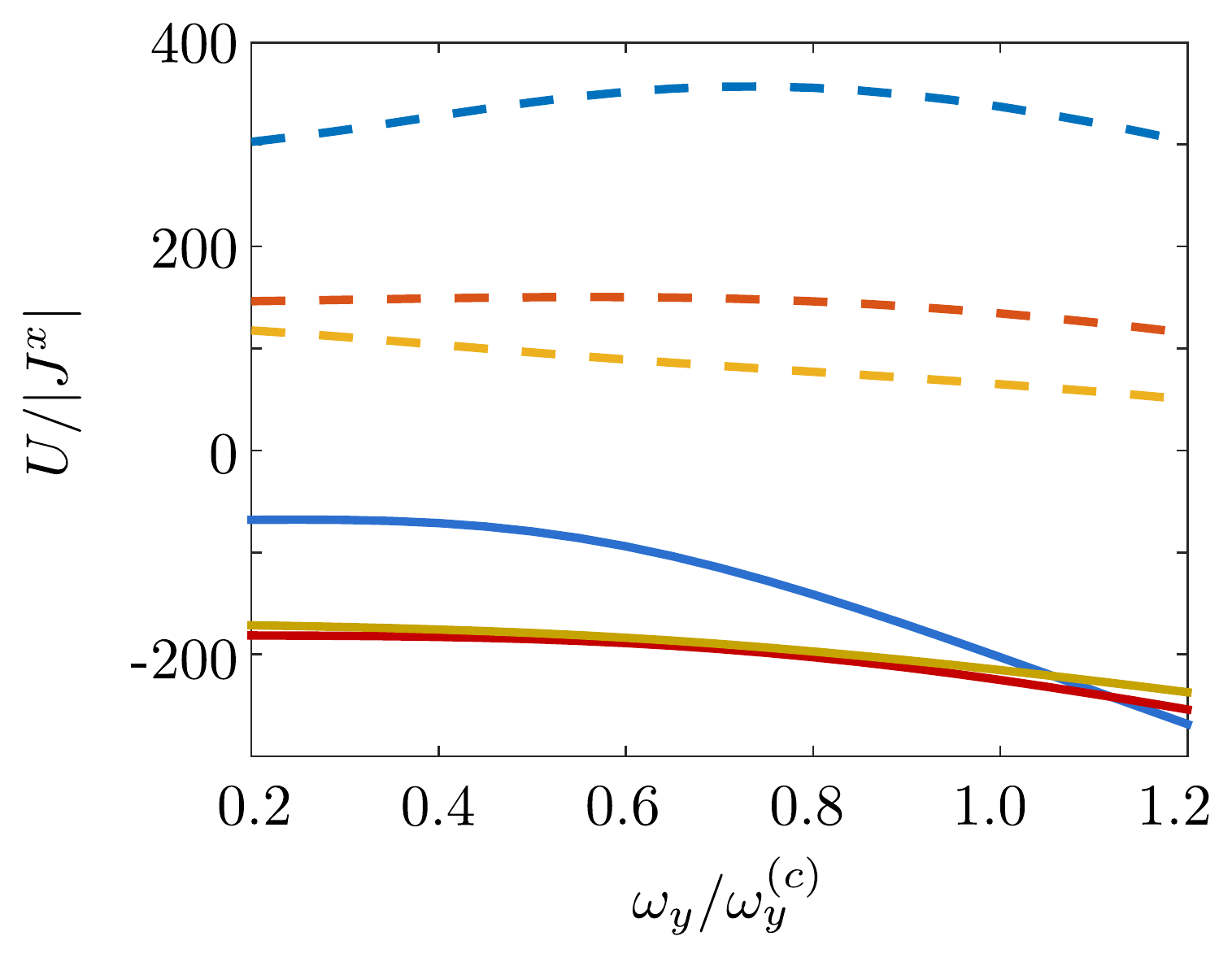}
		}
		\subfloat[]{
			\includegraphics[width=0.32\textwidth]{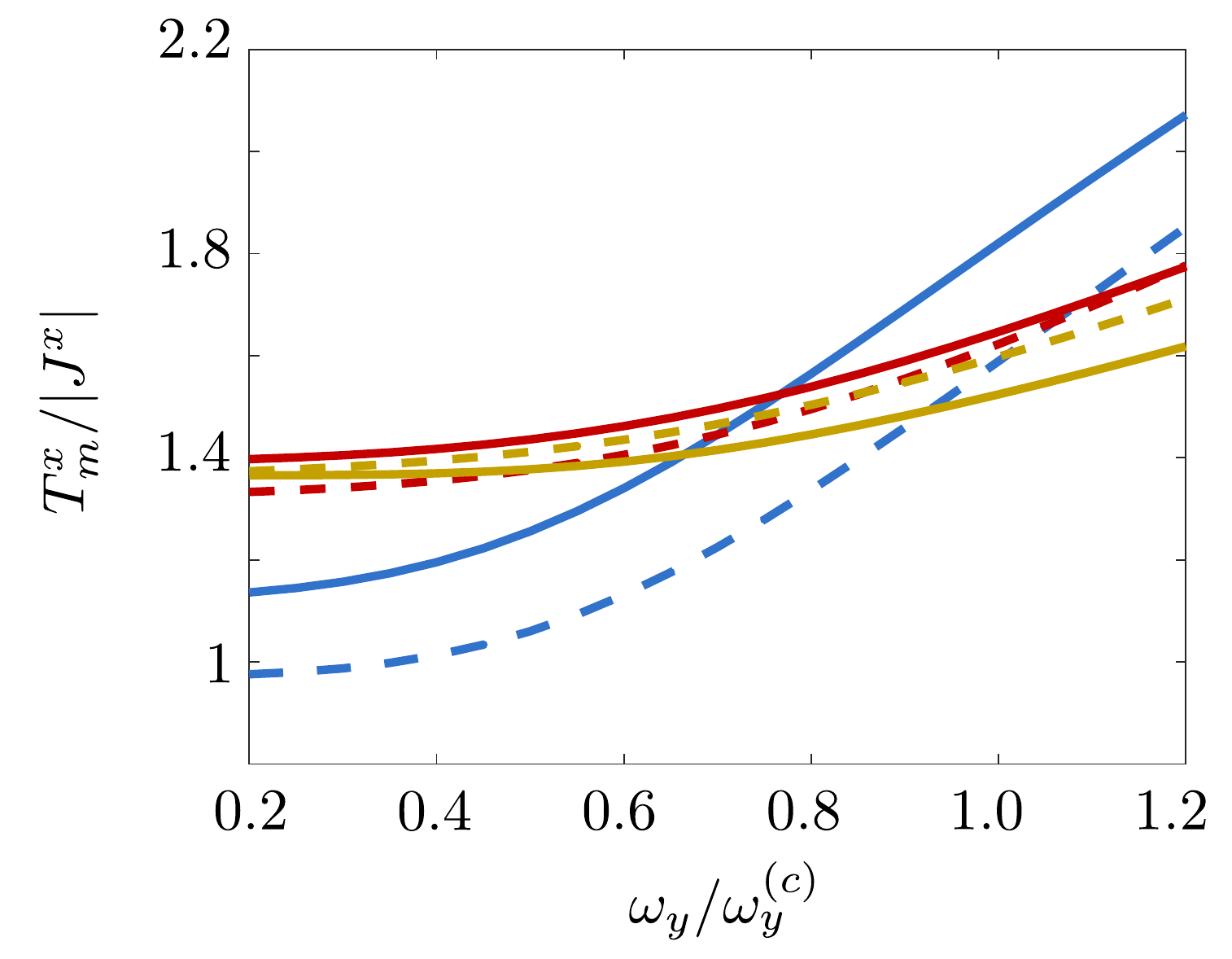}
		}
		\subfloat[]{
			\includegraphics[width=0.32\textwidth]{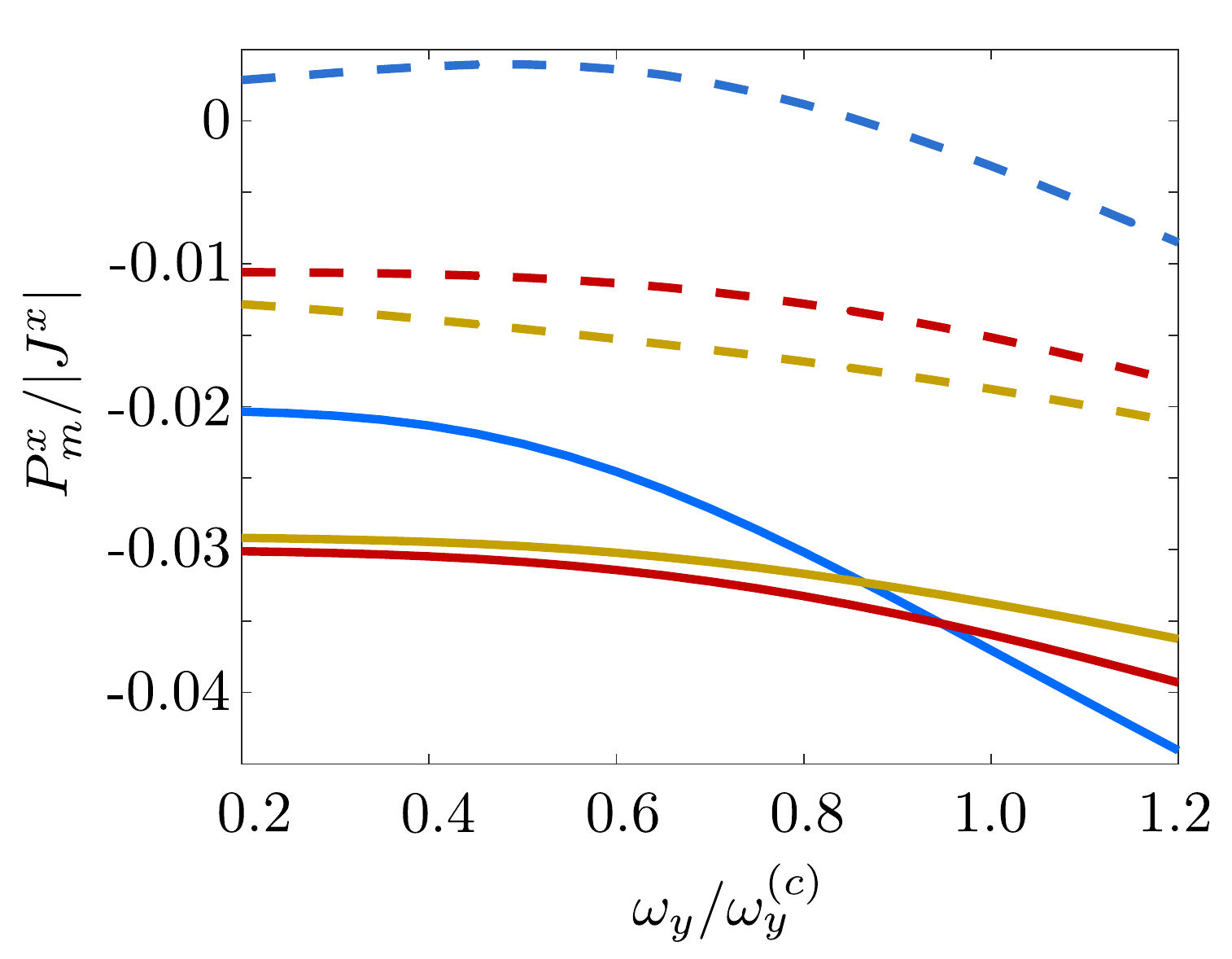}
		}
\caption{\label{Fig:2}(color online)  Coefficients of the multi-mode EBH Hamiltonian are reported as a function of $\omega_y$ (in units of $\omega_y^{(c)}$) for different values of $\sigma_z$:  (a) reports the on-site interaction coefficients $U^x_m$, (b) The density-assisted tunneling terms $T^x_m$ and (c) the pair tunneling terms $P^x_m$ (in units of $|J^x|$) for $\sigma_z=0.3375a$ (solid lines) and $\sigma_z=0.18a$ (dashed lines) and 
$m=0$ (blue line), $m=1$ (red line), 	$m=2$ (yellow line). The optical lattice depth and period are $V_L = 10 E_R$ (with $E_R$ the recoil energy) and $a=395$ nm, the scattering length is $a_S=a/50$,  the dipole moment is $p=1.15$ Debye, consistent with $^{85}$Rb-$^{133}$Cs bosonic molecules. }
\end{figure*}

\subsubsection{Determination of the Bose-Hubbard coefficients}

The coefficients corresponding to the interaction terms in the EBH model (Eqs. \eqref{BH:coeff:x1}-\eqref{BH:coeff:x4}, \eqref{BH:coeff:y}, \eqref{BH:coeff:xy}) explicitly depend on the confinement in the $z$ direction, which enters through the wave function $\theta_0(z)$ and specifically through the size of the wave function $\sigma_z$. We perform the integrals first analitically, by integrating out the $z$ variable  in Fourier space, then numerically. The details are reported in the Appendix. All other coefficients, which involve the integrals over two variables, are evaluated numerically. 

The dependence on the size of the trap along $z$, where the dipolar interaction is attractive, turns out to be relevant for certain parameter regimes, even if the motion is confined in the orthogonal plane \cite{Sowinski,Recati}. In Fig. \ref{Fig:2}(a) we can observe that increasing the size of the quantum fluctuations along $z$ can change the on-site interaction from being repulsive (positive coefficient) to become attractive (negative coefficient). Figures \ref{Fig:2}(b) and (c) show that varying $\sigma_z$ can substantially modify the strength of the density-assisted tunneling and of the pair tunneling terms, respectively. These results, moreover, highlight that there is an important interplay between the fluctuations along $y$ and $z$ which significantly affects the behaviour of the coefficients of $H^x$, and thus could change the phase of a quasi-one dimensional system of dipolar bosons.

\section{Quantum phases of small systems}
\label{Sec:2}

We now test the predictions of the multi-mode EBH model we derived by determining the quantum ground state as a function of the various parameters, as specified below. For this purpose we use exact diagonalization and assume periodic boundary conditions along $x$. This procedure limits us to small system sizes, yet it allows us to gain some insight into the possible phases one can observe. Moreover, it allows us to verify that our model reproduces correctly limiting cases analysed in the literature. This also provides us a point of comparison for future more elaborated numerical analysis based on Density Matrix Renormalization Group \cite{Silvi}. In this work we are specifically interested in determining the phase diagram as a function of (i) the depth of the optical lattice $V_L$, (ii) the $s$-wave scattering length, (iii) the transverse frequency $\omega_y$, (iv) the strength of the dipole-dipole interactions, (v) the size of the fluctuations along $z$. In this section we discuss the observables, which permit us to identify the quantum phases, and determine the phase diagrams in several limiting cases.

\subsection{Observables}	
	
	For a system of few sites we identify whether a phase is compressible by means of the local compressibility $\Delta n_j $, which is the expectation value over the ground state of the observable $\delta n_{j}$ and reads
	\begin{eqnarray*}
		\Delta n_j = \langle \delta n_j\rangle\,,\\
	\end{eqnarray*}
For a single mode EBH model, $\delta n_j=\delta n_j^0$ with $\delta n_j^0=n_j-\langle n_j\rangle$, and a phase is classified as incompressible when $\Delta n_j$ vanishes at all sites $j$. For our multi-mode EBH model we use
$\delta n_j=\delta n_j^M$, where 
\begin{equation}
\delta n_j^M=\sum_m (n_{j,m}- \langle n_{j,m}\rangle)^2\,.
\end{equation}
According to this criterion a phase is incompressible when $\Delta n_j=0$ at all sites $j$, like in the single-band case. 

Off-diagonal order is revealed by the non-vanishing value of the off-diagonal correlations (one-particle correlation function) $\phi$, which we define for the multi-mode EBH model as:
	\begin{eqnarray*}
		\phi=\sum_{j,m}\langle a^\dagger_{j,m} (a_{j+1,m} +a_{j-1,m})\rangle\,.
		\label{eq:phi}
	\end{eqnarray*}

The dipole blockade, scaling with coefficient $V_m^x$, favours the formation of a density modulation along $x$, which is signaled by a non vanishing value of the static structure form factor $S_x(q_x)$ at wave number  $q_x=\pi/a$. For the multimode EBH model we 
consider the structure form factor
	\begin{eqnarray}
	\label{eq:Sx}
	S_x(q_x)=\frac{1}{N^2}\sum_{j,l=1}^N{\rm e}^{\imath(j-l)q_xa} \sum_m\left(\langle n_{j,m}n_{l,m}\rangle-\langle n_{j,m}\rangle\langle n_{l,m}\rangle\right)\,.
	\end{eqnarray}
The value $S_x(\pi/a)\neq 0$ signals the formation of a structural order. We denote the phase by super-solid (SS) when this occurs in a compressible phase with non-vanishing off-diagonal correlations. The phase is instead charge-density wave (CDW) when incompressible \cite{Goral,Sowinski}. 

Additionally, pair tunnelling terms are expected to favour the onset of what has been denoted by pair superfluidity \cite{Sowinski}, and which shall be signaled by a non-vanishing expectation value of the pair-correlation function, defined as:
\begin{equation}
\label{eq:Phi}
\Phi = \sum_{j,m}\langle a^\dagger _{j,m} a^\dagger_{j,m} a_{j+1,m} a_{j+1,m} + {\rm H.c.}\rangle\,.
\end{equation}

\begin{figure*}
		\centering
		\subfloat{
			\includegraphics[width=0.32\textwidth]{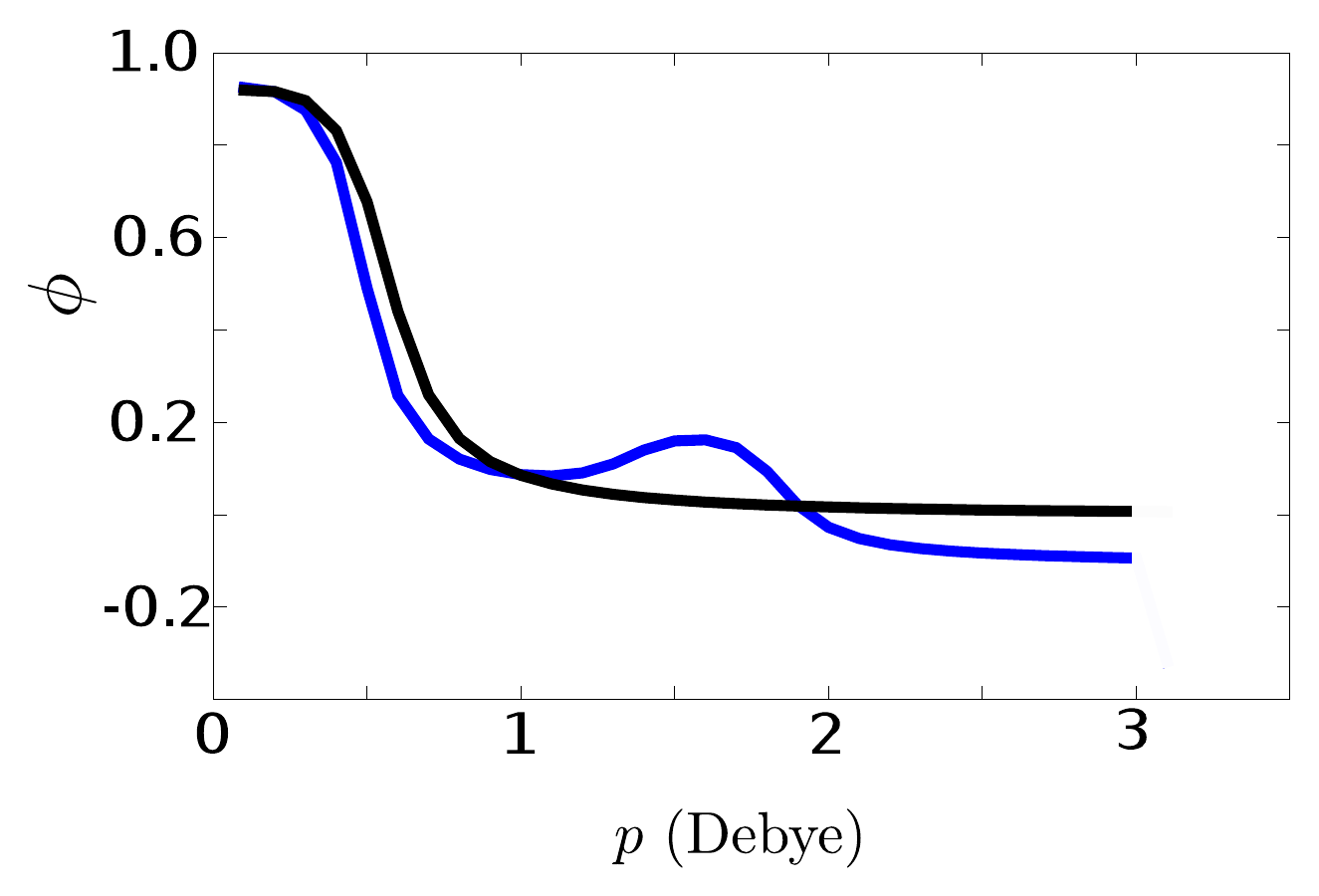}
		}
		\subfloat{
			\includegraphics[width=0.32\textwidth]{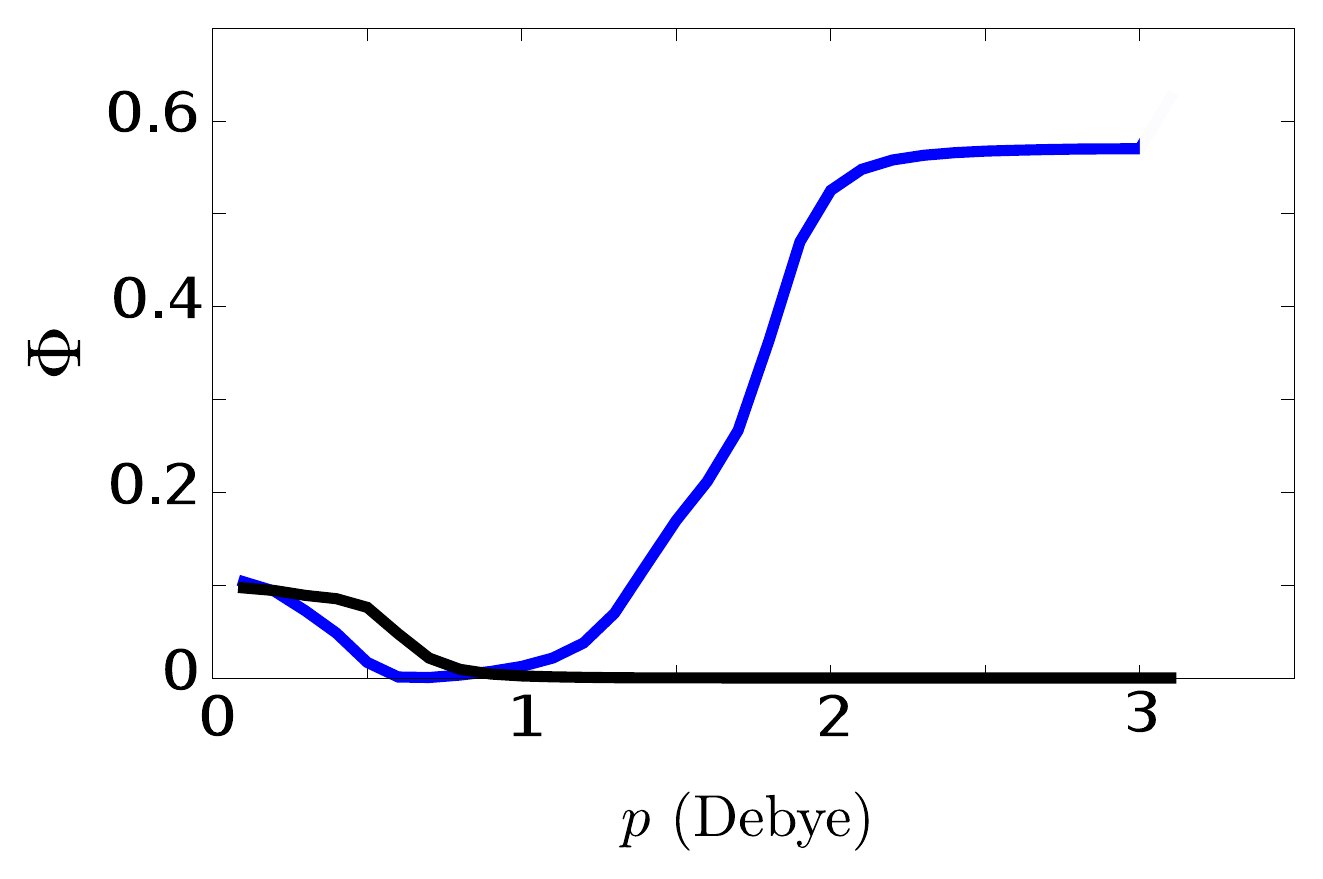}
		}
		\subfloat{
			\includegraphics[width=0.32\textwidth]{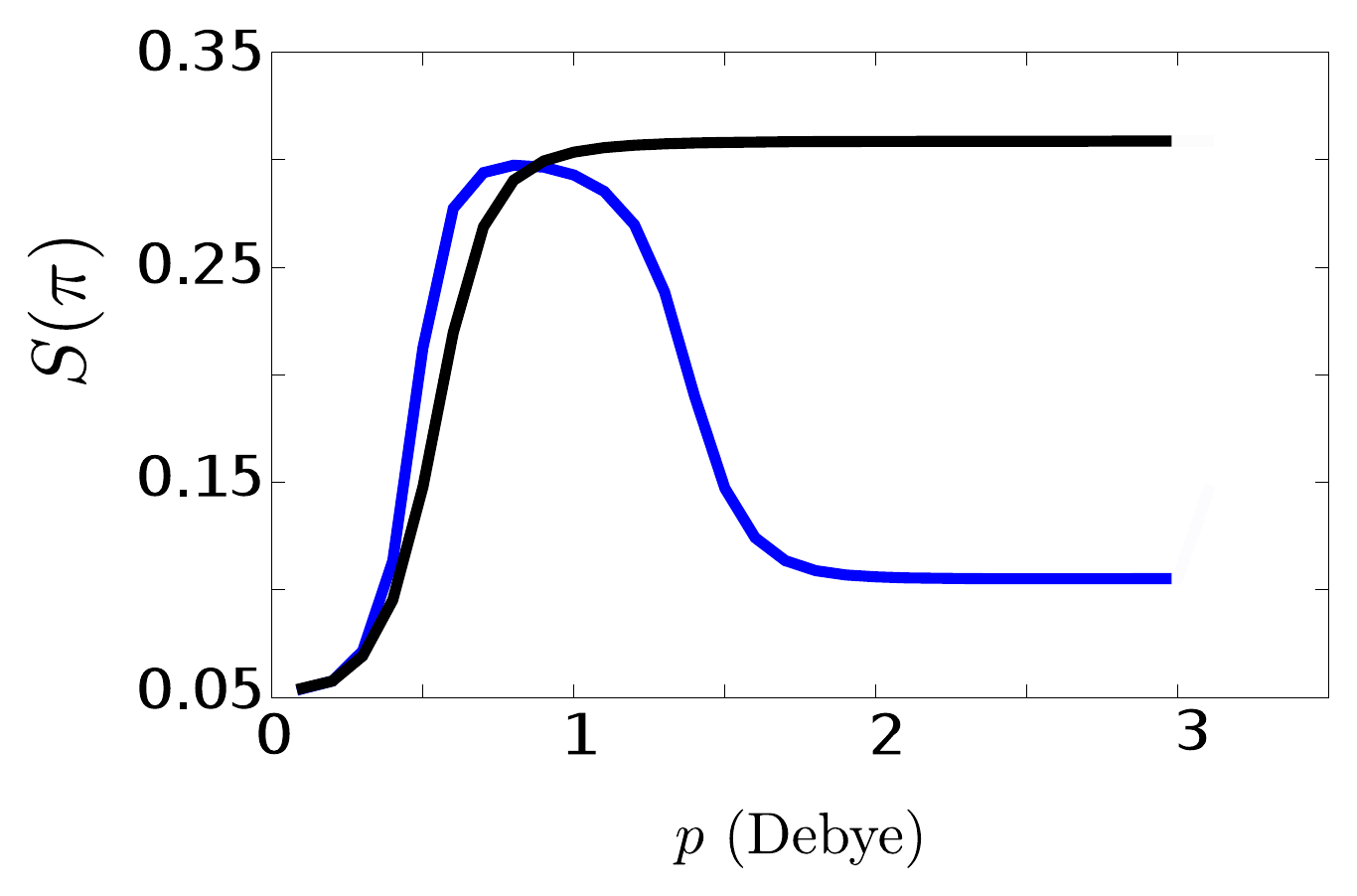}
		}
		\caption{\label{Fig:Tagliacozzo} (color online) Real part of (a) single-particle correlation $\phi$, Eq. \eqref{eq:phi}, (b) two-particle correlations $\Phi$, Eq. \eqref{eq:Phi}, and (c) $S(\pi)\equiv S_x(\pi/a)$, Eq. \eqref{eq:Sx}, as a function of the dipole moment $p$ (in Debye) for a lattice of 12 sites filled with 6 particles and periodic boundary conditions. The blue curve is obtained for the ground state of Hamiltonian $H^x_0$, Eq. \eqref{H:x}, the black curve is found when we arbitrarily set $T^x=P^x=0$ in Hamiltonian $H^x_0$. The parameters are $V_L=6E_R$, $a_S=a/100$, and $\sigma_z = 0.2279 \, a$. The other parameters are the same as in Fig. \ref{Fig:2}. Note that for each value of $p$ we modify the trap frequency $\omega_y$ according to the prescription $\int dx\, x^2 w^2(x) = \int dy\, y^2 \phi_0^2(y)$, see text. }
\end{figure*}

These quantities have been used in the literature to characterize the phases of the one-dimensional EBH model, their expectation value varies with the strength of the dipolar moment, as summarized in Fig. \ref{Fig:Tagliacozzo}, which reproduces the behaviour reported in Ref. \cite{Sowinski}. For completeness, we mention that the one-dimensional EBH model can also exhibit a topological phase, denoted by Haldane-insulator phase, which is incompressible and characterised by $S_x(\pi/a)=0$ \cite{Batrouni2013}. The so-called string-order operator $O_s(|j-l|)$ signals its appearance \cite{Berg,Batrouni2013,Batrouni2014}. We will omit to analyse its expectation value for the small system sizes we consider, since a non-vanishing expectation value is not meaningful.
 	
In addition to this set of observables, we also consider the structure form factor at wave number $q_y=\pi/a$, which signals the onset of zigzag order and is defined as: 
\begin{equation}
	\label{S:zigzag}
S_y \left(\frac{\pi}{a}\right)= \frac{1}{N^2} \sum_{j\neq l} (-1)^{j-l} \langle y_j y_l\rangle \\
\end{equation}
where 
	\begin{equation}
	y_j = \sum_{m,n} Y_{mn} \,a_{j,m}^\dagger a_{j,n}\,,
	\end{equation}
and $$Y_{m,n} = \int dy\, y \,\phi_m^*(y) \phi_n(y)$$ is a real matrix whose elements depend on the physical parameters \cite{Silvi}. 
When $S_y (\pi/a)\neq 0$, the dipoles form a zigzag transverse structure.

\subsection{Phase diagrams}
	\label{sec:phase-diagram}
	
We now report phase diagrams for the salient properties of Hamiltonian $H_{BH}$, Eq. \eqref{H:BH}, evaluated by means of exact diagonalization on a lattice with periodic boundary conditions along $x$ and composed of 4 to 12 sites. The considered number of sites for a given phase diagram depends on the number of transverse modes we need to take into account in order to warrant the convergence of the calculations. In what follows we use the parameters of $^{85}$Rb-$^{133}$Cs bosonic molecules with electric dipole moment of $p_0=1.15$ Debye \cite{Naegerl}, confined by an optical lattice along $x$ at the interparticle distance $a=395$ nm, corresponding to half wavelength of the standing-wave laser, unless otherwise stated. The parameters $a$ and $p_0$, moreover, are the units of length and of the dipole moment we will refer to.  

\subsubsection{Phase diagram of the quasi one-dimensional array}
	\label{sec:quasi-1d}

We first consider the limit in which the trapping frequency $\omega_y>\omega_y^{(c)}$, so that the transverse motion is in the ground state of the transverse oscillator and the model is reduced to a single-band EBH model, described by Hamiltonian $H^x_0$, corresponding to Eq. \eqref{H:x} with $m=0$. We are interested first in reproducing the results of  Ref. \cite{Sowinski}  with our multi-mode EBH model and therefore need to identify the conditions on the trap frequency $\omega_y$ for which we reproduce the single- and two-particle correlations and the component of structure form factor $S_x(\pi/a)$ as a function of $p$, when we fix the other corresponding parameters. For each value of $p$ we choose $\omega_y$ such that the width of the lowest eigenfunction $\phi_0(y)$ is equal to the width of the Wannier functions for the given lattice depth, $\int dx\, x^2 w^2(x) = \int dy\, y^2 \phi_0^2(y)$. The inequality $\omega_y>\omega_y^{(c)}$ is fulfilled for $p<3$ Debye. 

Figure \ref{Fig:Tagliacozzo} displays $\phi$, $\Phi$, and $S_x(\pi/a)$ as a function of the strength of the dipole moment $p$, in units of $p_0$, for 12 sites and at half-filling. The results reproduce the behaviour reported in Ref. \cite{Sowinski}. In order to highlight the role of pair and density dependent tunneling, in all figures we also give the value obtained by setting $T^x=P^x=0$. This comparison shows that these terms are essential for the appearance of two-particle correlations, signaling pair superfluidity. This occurs at sufficiently large value of $p$, which in turn scales the corresponding coefficients $T^x$ and $P^x$. 

\begin{figure}
		\centering
		\subfloat[ \label{fig:spc_g2}]{
			\includegraphics[width=0.48\textwidth]{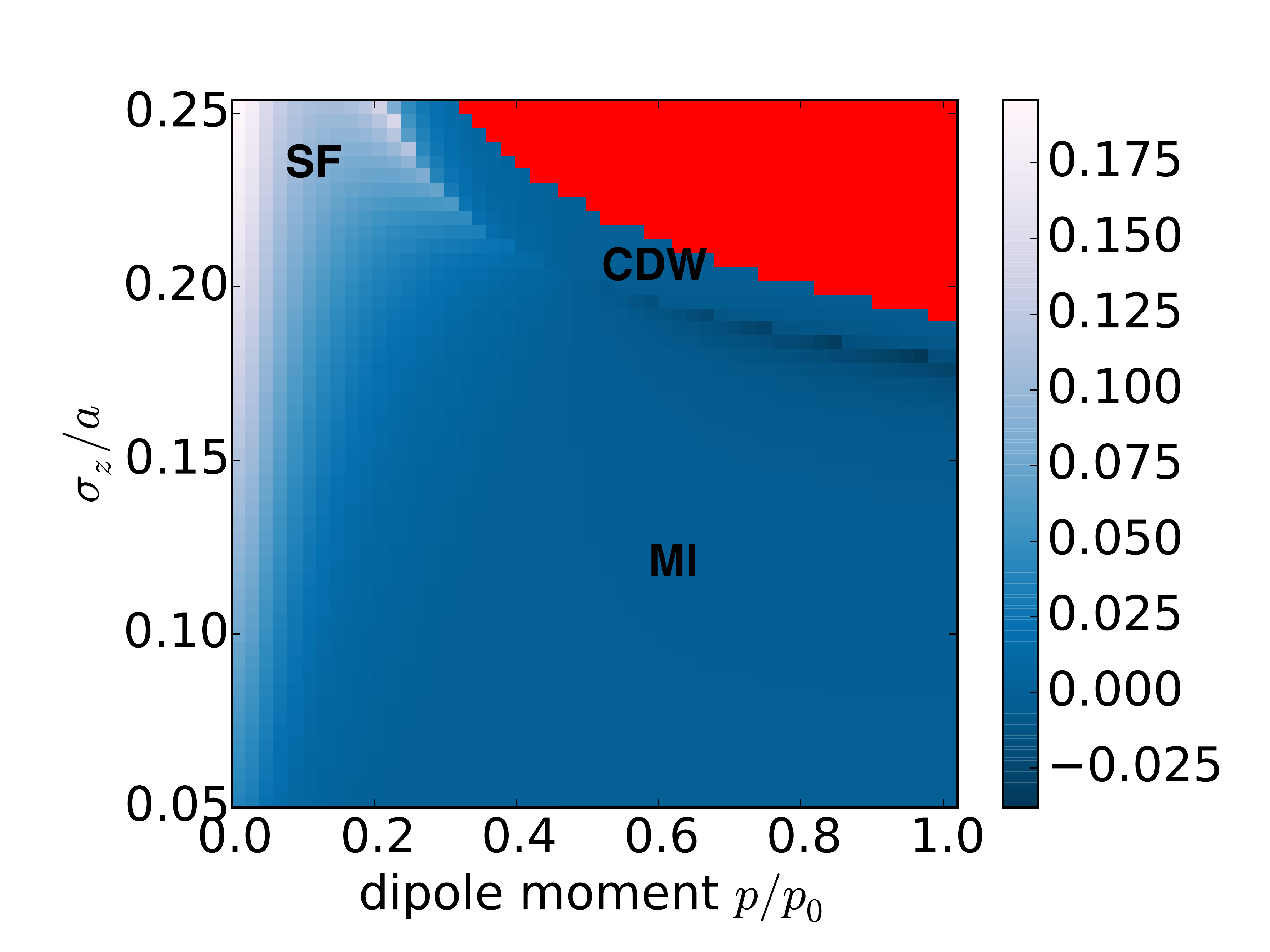}
		}\\
		\subfloat[ \label{fig:tpc_g2} ]{
			\includegraphics[width=0.45\textwidth]{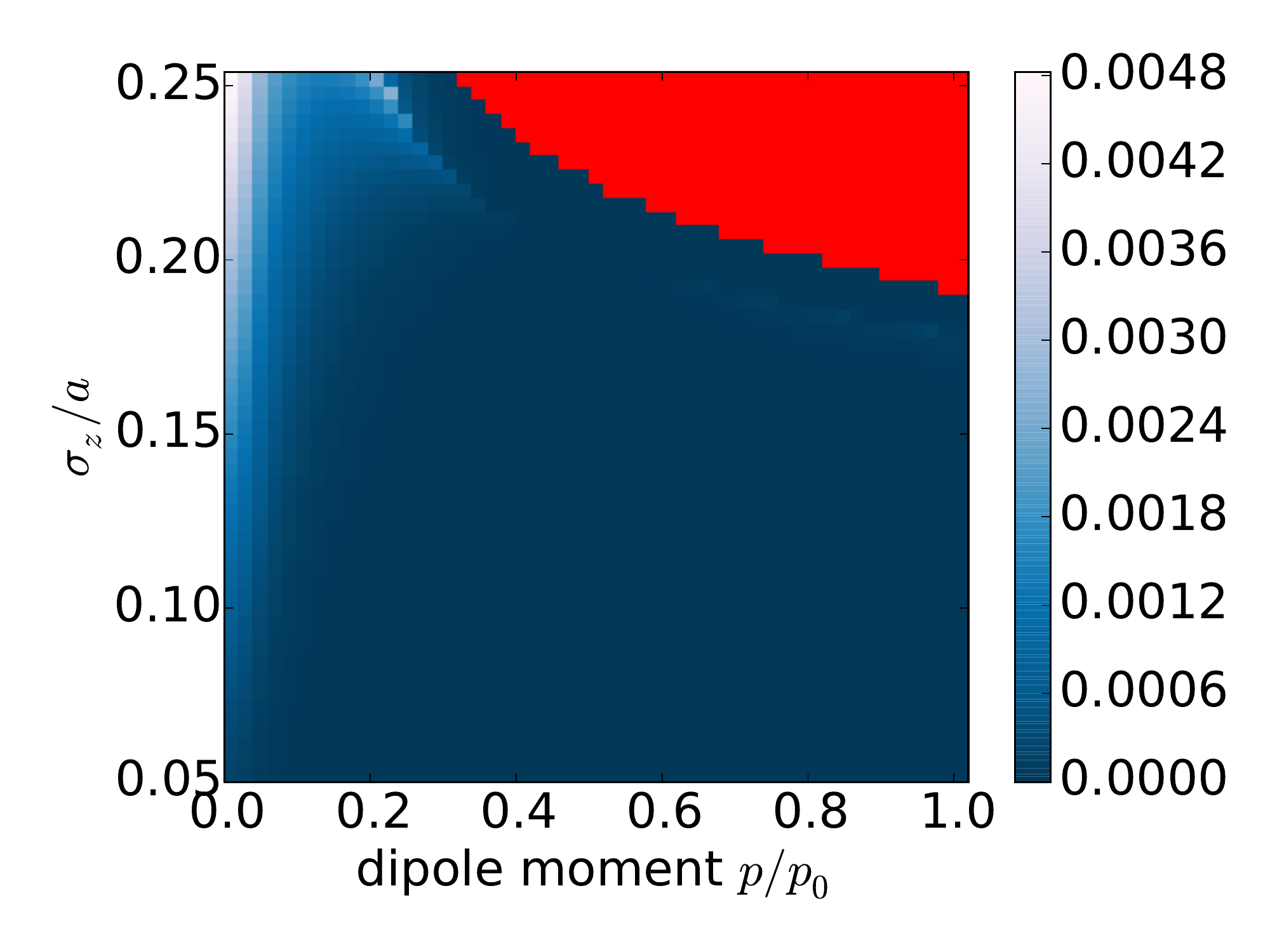} 
		}\\
		\subfloat[\label{fig:st_g2}]{
			\includegraphics[width=0.45\textwidth]{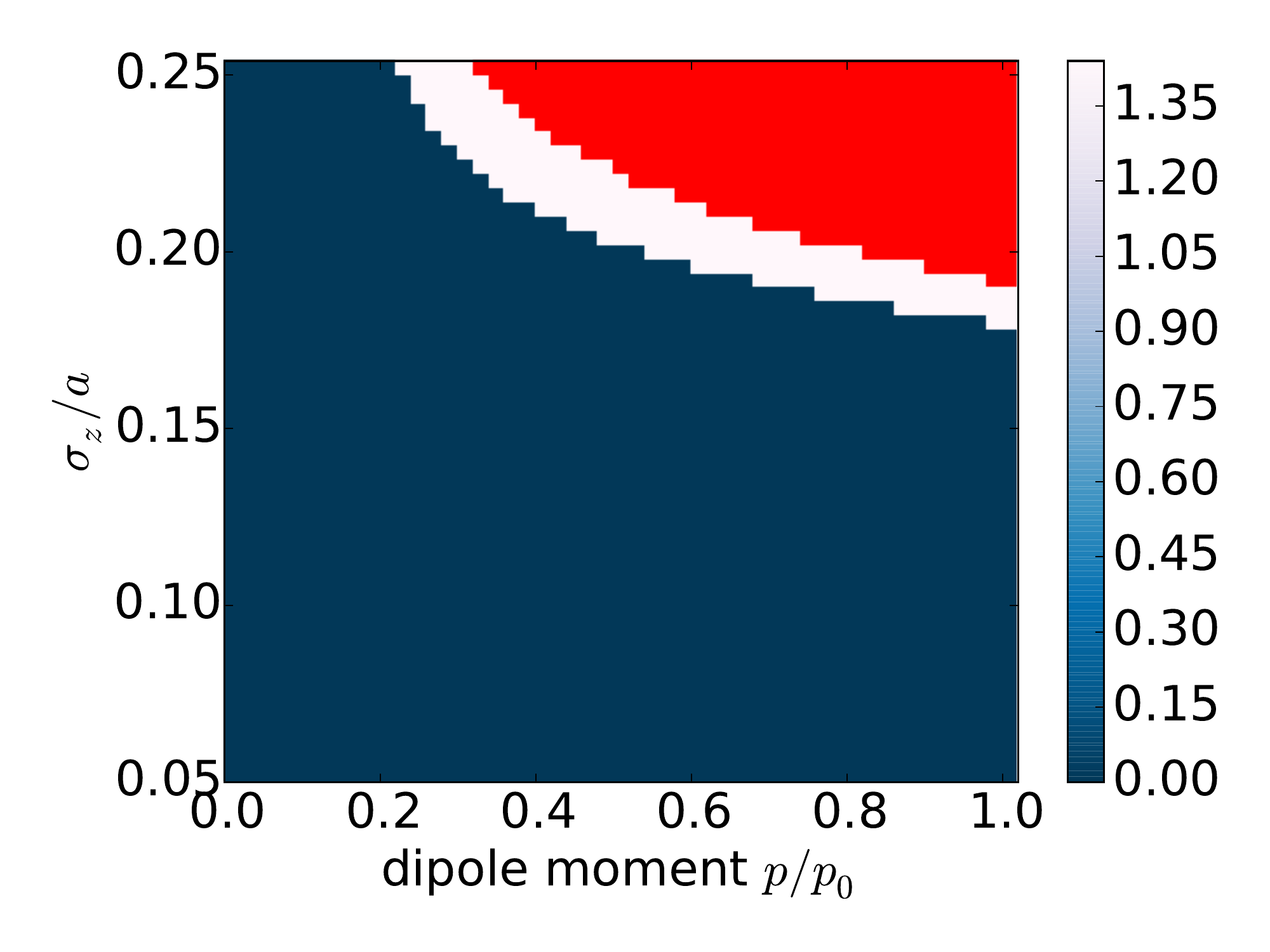}
		}
		\caption{\label{Fig:4}(color online) Contour plot of the real part of the single particle correlations $\phi$ (a), the two particle correlations $\Phi$ (b) and the structure form factor $S_x(\pi/a)$ (c) as a function of the width $\sigma_z$ (in units of $a$) and of the dipole moment $p$ (in units of $p_0=1.15$ Debye) for unit filling. The red area denotes the region where the on-site interaction $U_0^x$ is negative. The parameters are $V_L = 20 E_R$, $\omega_y = 1.45 \omega_y^{(c)}$,  and $a_s=a/50$. }
	\end{figure}

We now extend this analysis to unit fillings, $\Braket{n_j} = 1$. This regime was not considered in Ref. \cite{Sowinski} since the role of pair superfluidity is expected to be small, nevertheless it is relevant for studying the linear-zigzag transition. In order to benchmark this case, we determine the phase diagram in the limit where the lowest transverse band is occupied. Figures \ref{fig:spc_g2}-\ref{fig:st_g2} display the contour plots of the single particle correlation $\phi$, the two particle correlation $\Phi$ and the structure factor $S_x(\pi/a)$, respectively, as a function of the steepness of the confinement along the $z$-axis, $\sigma_z$, and of the strength of the dipole moment for the ground state of a lattice composed by 10 sites and 10 particles. The red-coloured region indicates the unstable regime, where the on-site interaction coefficient $U_0^x$ becomes negative: The border of this region is the line where $U_0^x=0$. Close to the unstable area, at small values of $U_0^x$  there is a striped region where the single and two particle correlations, $\phi$ and $\Phi$, vanish. We verified that the local compressibility also vanishes. In the same parameter area, the structure factor $S_x(\pi/a)$ is different from zero, see (c); we hence conjecture that the system is in a CDW phase. This conjecture is further supported by the exponential decay of the long range correlations $\phi_{l} = \langle a_j^\dagger a_{j+l}\rangle$. Outside of this region, in the stable regime and for $p$ sufficiently large the phase is characterized by an exponential decay of $\phi_l$ and by $S_x(\pi/a)=0$. We identify it with a MI phase. At small values of $p$ as well as close to the border separating with the CDW phase the system is SF. We do not find signatures indicating a  PSF phase in the parameter regime we explored: $\Phi$ is different from zero in the region where $\phi\neq 0$ and scales as in a SF phase. 

The peculiarity of these results can be better highlighted by considering the phase of the system as a function of $p$ at fixed values of $\sigma_z$. For sufficiently small values of $\sigma_z$, by increasing $p$ we observe a transition from SF to MI. At sufficiently large value of $\sigma_z$ increasing $p$ leads to a transition from SF to a CDW, before the system becomes unstable. Between these two regimes, there seem to be a small interval of values $\sigma_z$ where the system goes from SF to MI to SF by increasing $p$. 

\begin{figure}
		\centering
		\subfloat[\label{fig:spc_p1}]{
			\includegraphics[width=0.48\textwidth]{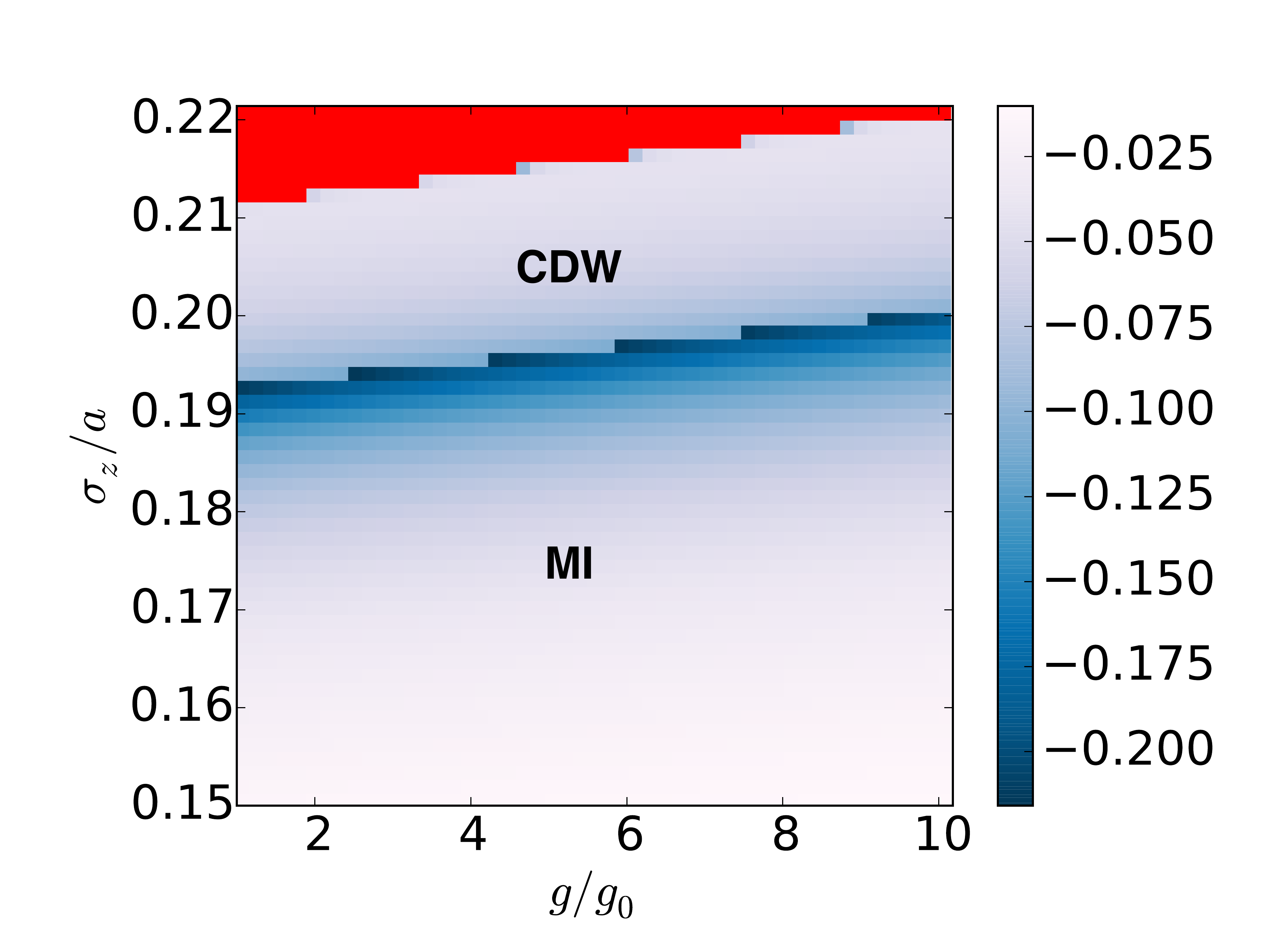}
		}\\
		\subfloat[\label{fig:tpc_p1}]{
			\includegraphics[width=0.45\textwidth]{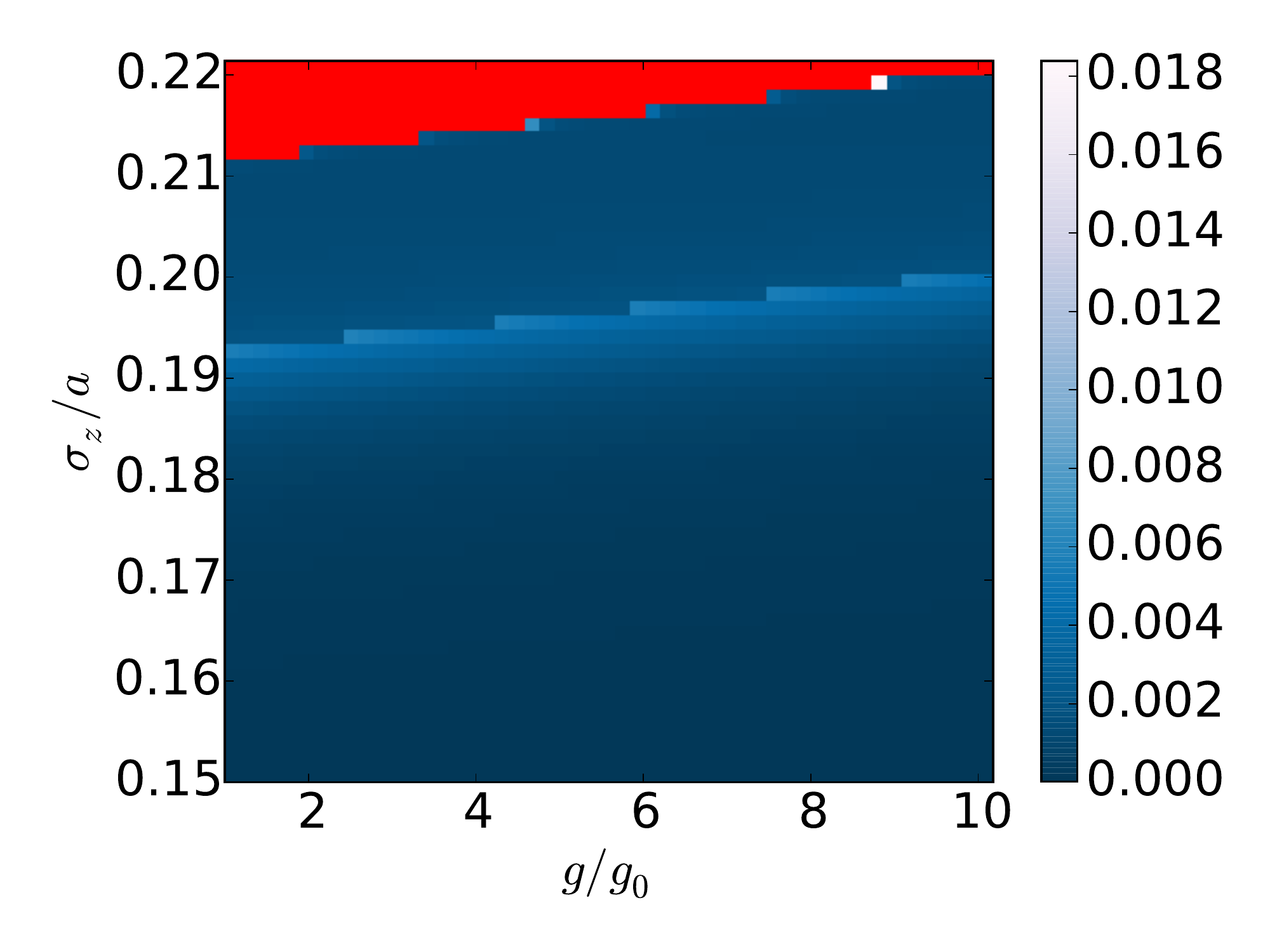}
		}\\
		\subfloat[\label{fig:st_p1}]{
			\includegraphics[width=0.45\textwidth]{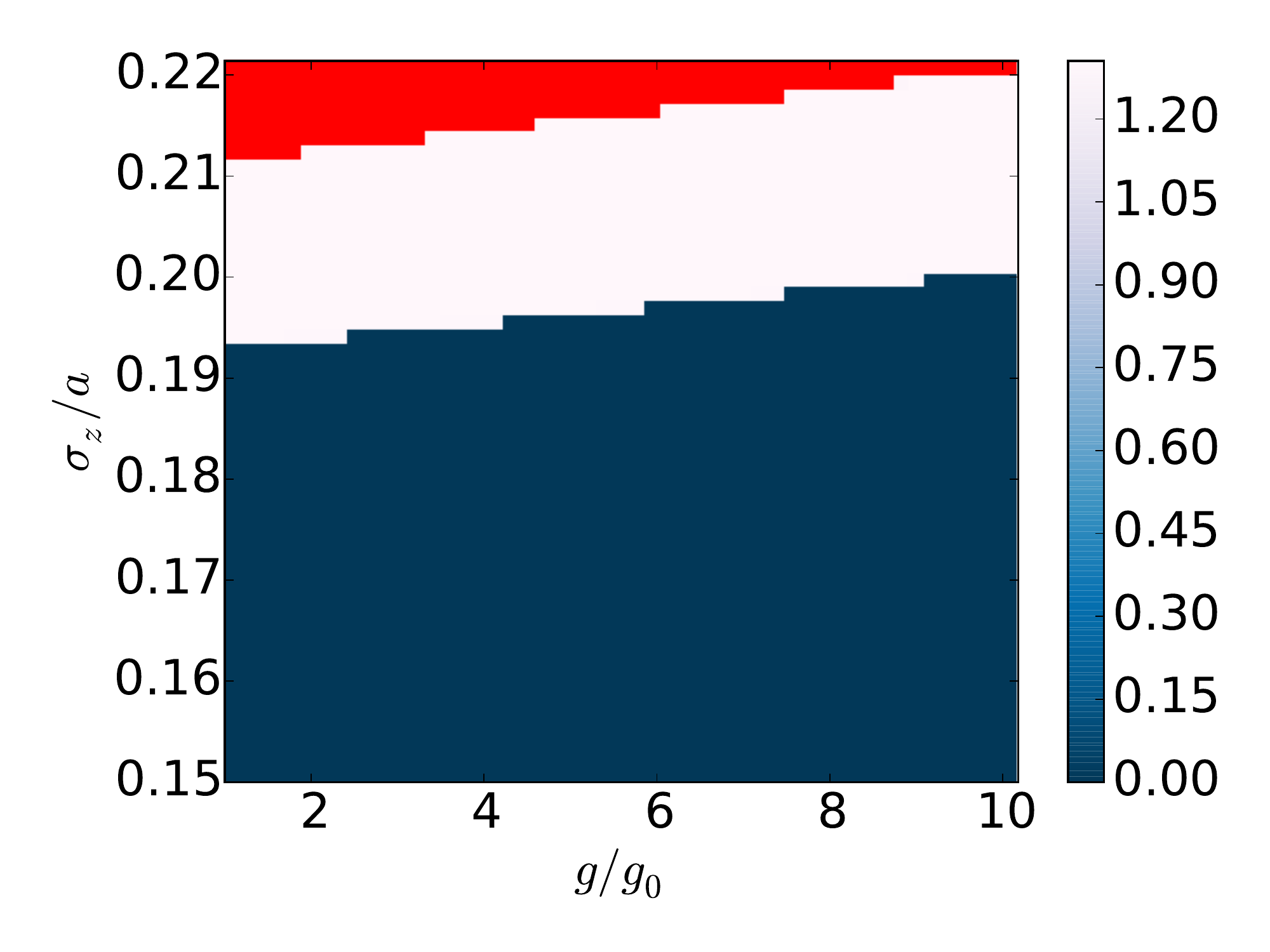}
		}
		\caption{\label{Fig:5}(color online) 
		Contour plot of the real part of the single particle correlations $\phi$ (a), the two particle correlations $\Phi$ (b) and the structure form factor $S_x(\pi/a)$ (c) as a function of the width $\sigma_z$ (in units of $a$) and of the onsite scattering strength $g$ (in units of $g_0=4\pi a_s/M$ with $a_s =a/50$) and for unit filling. The red area denotes the region where the on-site interaction $U_0^x$ is negative. The parameters are $V_L = 20 E_R$, $\omega_y = 1.45 \omega_y^{(c)}$, and $p=1.15$  Debye. }	
	\end{figure}
Figures \ref{fig:spc_p1}-\ref{fig:st_p1} show the behaviour of $\phi$, $\Phi$ and $S(\pi)$ when varying the strength of the onsite interaction while keeping $p$ constant. The behaviour reported in these plots can be put in direct connection with the ones of Fig. \ref{Fig:4} since increasing $g$ partly corresponds to effectively decreasing $p$. Here, we clearly observe that CDW and MI are separated by a discontinuity in the structure form factor and in the single particle correlations, which occurs at the same value of $\sigma_z$.
In order to determine the properties at the discontinuity we calculated the susceptibility of the ground-state fidelity  $\mathcal{F}(\sigma_z) = |\Braket{\Psi(\sigma_z) | \Psi(\sigma_z  + \delta)}|$ , defined as \cite{Fidelity}
$$ \chi = \left. \frac{\partial^2 \mathcal{F}(\sigma_z)}{\partial \delta^2}  \right|_{\delta \to 0}\,.$$
The susceptibility is different from zero at the point where the structure factor exhibits a discontinuity. We verified that its value increases with the particle numbers. On this basis we conjecture that this discontinuity signals a quantum phase transition. 

\subsubsection{The multi-mode model at the linear-zigzag instability}

\begin{figure}
	\includegraphics[width=0.5\textwidth]{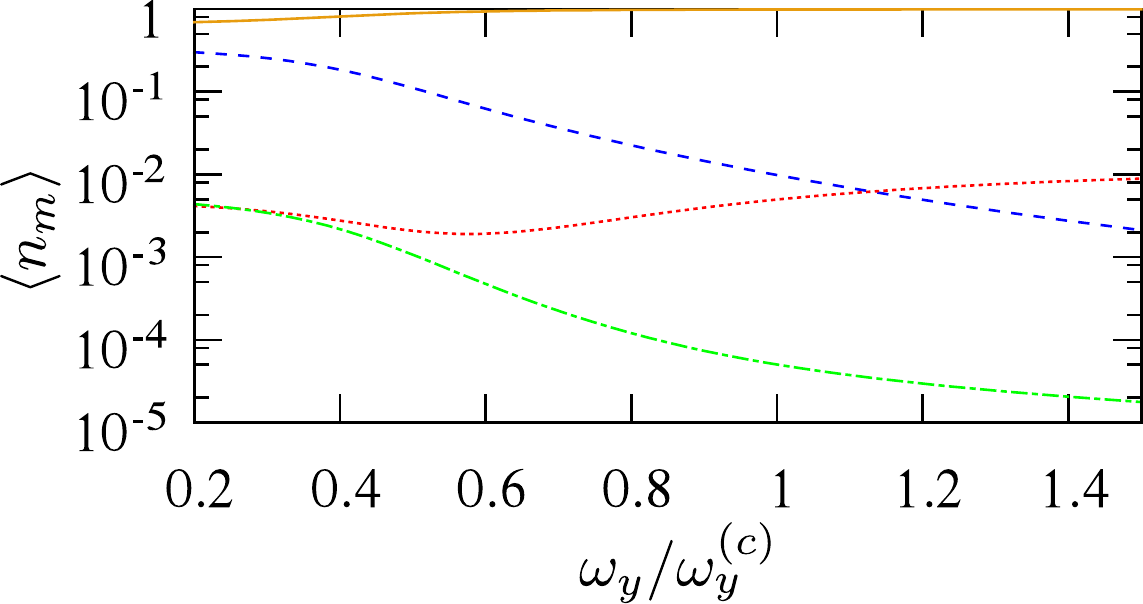}
	\caption{(color online) Occupation $n_{m} =  \braket{n_{j,m}}$ of the local Hamiltonian eigenstate $m$ as a function of the trap frequency $\omega_y$ for $m=0$ (orange line), $m=1$ (red dotted), $m=2$ (blue dashed) and $m=3$ (green dashed-dotted). The calculation has been performed for $N=6$ particles over 6 sites. The parameters are $V_L = 6 E_R$, $\sigma_z = 0.1a$, $g=g_0$ and $p=p_0$. \label{Fig:6}}
\end{figure}

We now report properties of our model at the linear-zigzag instability and for unit filling. Including extra bands is here necessary but it severely limits the computational capability of exact diagonalization, since the dimensionality of the problem rapidly scales up with the number of orbitals. We first check how many states of the local basis shall be considered in order to warrant the convergence of the calculations for $\omega_y\sim \omega_y^{(c)}$. Figure \ref{Fig:6} displays the occupation of the lowest four orbitals ($m=0,1,2,3$) as a function of $\omega_y$ for $N=6$ particles in a relatively shallow optical lattice, $V_L = 6 E_R$. For $\sigma_z = 0.1a$ and for the considered values of the trap frequencies  $\omega_y > \omega_y^{(c)}$ we find that only the lowest orbital is relevant, while  for  $\omega_y < \omega_y^{(c)}$,  also the second orbital is occupied. Recalling that first orbital is even and the second orbital is odd,  this change of occupation corresponds to the onset of the zigzag phase.
In both cases, 99 \% of the population is in the lowest two bands.  This result remarkably shows that the basis decomposition we perform warrants a fast convergence even at transverse trap frequencies well below the mean-field critical value.

\begin{figure}
	\includegraphics[width=0.5\textwidth]{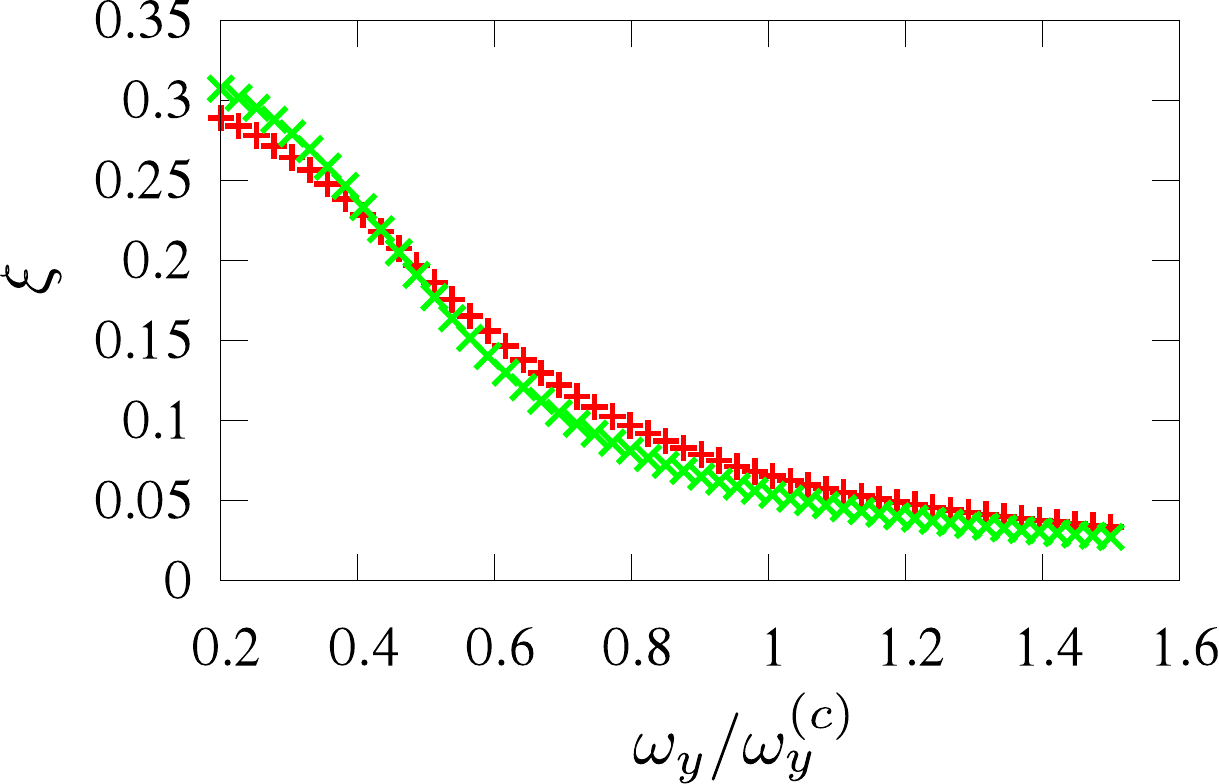}
	\caption{(color online) The structure factor $\xi$, Eq. \eqref{S:zigzag}, indicating zigzag order for $\sigma_z=0.1\,a$ at an optical lattice depth of $V_L=15 E_R$ for unit filling with 4 sites (red points) and 6 sites (green) as a function of $\omega_y / \omega_y^{(c)}$. The calculations were done with exact diagonalization using the first 4 orbitals. The other parameters are the same as in Fig. \ref{Fig:6}.
		\label{Fig:7}}
\end{figure}

Figure \ref{Fig:7} shows the zigzag order parameter $\xi=S_y(\pi/a)$, Eq. \eqref{S:zigzag}, as a function of $\omega_y$ for $N=4,6$ and a steeper optical lattice: $\xi$ increases by decreasing $\omega_y$, allowing to identify the zigzag phase. The intersection between the two curves at $N=4$ and $N=6$ suggests the location of the critical point, which is at a smaller value than the mean-field prediction and consistent with the DMRG result of Ref.~\cite{Silvi}.

\section{Conclusions}
\label{Sec:Conclusions}

In this work we have derived a multimode EBH model which can naturally describe the effects of the quantum fluctuations of an array of dipoles at the structural transition to zigzag order. Our model takes into full account the three-dimensional, anisotropic nature of the dipolar interaction. Our results show that the frequency of the transverse confinement controls not only the onset of zigzag order, but also determines the quantum phases of the molecules along the chain. The interplay between classical and quantum effects as a function of the transverse confinement is an open question, which will be addressed in future works performing numerical simulations with large numbers of particles. For this purpose the study here presented provides an important benchmark. Moreover, our model could be extended to describe structural transitions of cold polar molecules in arrays of one-dimensional tubes, in the setup analysed in Refs. \cite{Kollath,Knap}.

\acknowledgments

The authors acknowledge discussions with Efrat Shimshoni, Andr\'e Winter, and Pietro Silvi. They are especially grateful to Rebecca Kraus for the critical reading of this manuscript. Financial support by the German Research Foundation (DFG, GiRyd  Priority Programme 1929 "Giant Interaction in Rydberg Systems") is gratefully acknowledged.

\appendix

\section{Determination of the Bose-Hubbard coefficients}

In this Appendix we derive the effective dipole-dipole interaction in two dimensions by integrating out the motion along the $z$-axis in the integrals needed to evaluate the coefficients for  Eqs. \eqref{H:x}, \eqref{H:y}, and \eqref{H:xy}. 
In order to illustrate the procedure we first write these terms in generic form as 
\begin{align}
\label{ToIntegrate}
{\mathcal V} = \int d\bm \rho  A(\bm \rho) \int d{\bm \rho'}V_{2D}(\bm \rho - {\bm \rho'}) B({\bm \rho'}),
\end{align}
where ${\bm \rho}=(x,y)$ and $V_{2D}(\bm \rho - {\bm \rho'})$ contains the integrals in the $z,z'$ variables and specifically takes the form:
\begin{eqnarray}
V_{2D}({\bm \rho} - {\bm \rho'}) &=& \int \, dz_1 dz_2 \, \theta_0^2(z_1) \theta_0^2(z_2) U_d(\bm r_1 - \bm r_2)\,.
\label{2D}
\end{eqnarray}
Thus, $V_{2D}({\bm \rho} - {\bm \rho'})$ is an effective dipole-dipole interaction in two dimensions. Its form can be simplified by 
using center-of-mass $Z=(z_1+z_2)/2$ and relative variable $z=z_1-z_2$. After integrating out the center-of-mass variable, Eq. \eqref{2D} reads
\begin{eqnarray}
V_{2D}(x,y) = \int dz \, \frac{1}{\sqrt{2 \pi} \sigma_z} e^{-z^2/(2 \sigma^2)} U_d(x,y,z)\,,
\label{2D:1}
\end{eqnarray}
where $x=x_1-x_2$ and $y=y_1-y_2$. We then determine the integral in Eq. \eqref{ToIntegrate} using a convolution method~\cite{Wall2013}. This consists first in writing the integral in Eq. \eqref{ToIntegrate} as
\begin{align}
{\mathcal V}  = \int d{\bm \rho} A({\bm \rho}) C({\bm \rho})\,,
\end{align}
where we dropped the indices for convenience and introduced $C({\bm \rho})$, which is defined as
\begin{align}
C({\bm \rho})=\mathcal{F}^{-1}_{\bm k}\left[ \mathcal{F}_{\bm k}\left[V_{2D}({\bm \rho})\right] \mathcal{F}_{\bm k}\left[B({\bm \rho})\right] \right]
\end{align}
with $\mathcal{F}_{\bm k}$ the Fourier transform in two dimensions (${\bm k}=(k_x,k_y)$) and $\mathcal{F}_{\bm k}^{-1}$ its inverse. This procedure allows one to calculate the integral by computing a 2D Fourier transform and a 2D integral, instead of integrating a four dimensional integral in real space, thus saving computing time and allowing one to use a finer grid of discretization. 

The Fourier transform $ \mathcal{F}_{\bm k}\left[V_{2D}({\bm \rho})\right]$ can be explicitly calculated. We first use the definition of the inverse Fourier transform:
\begin{align}
V_{2D}(x,y)&=\mathcal{F}^{-1}_{\bm k}[ \tilde V_{2D}({\bm k})]\nonumber\\
    		 &=\frac{1}{(2\pi)^2} \int d k_x \,d k_y \, e^{i (k_x x + k_y y)} \tilde V_{2D}({\bm k})\,,
		 \label{2D:2}
\end{align}
We further observe that Eq. \eqref{2D:1} can be rewritten as 
\begin{align}
V_{2D}(x,y) &= \int dz \, \int \frac{dk_z'}{2\pi} \, \tilde A(k_z') e^{i k_z' z} \nonumber\\
&\times \int   \frac{d^3k}{(2 \pi)^3}\, \tilde U(k_x,k_y,k_z) {\rm e}^{i (k_xx+k_yy+k_zz)}\,,
\label{2D:3}
\end{align}
where $\tilde A(k_z) = {\rm e}^{-k_z^2 \sigma_z^2 / 2}$ and
\begin{align}
\tilde U(k_x,k_y,k_z) &= \int d^3r \,  {\rm e}^{-i (k_xx+k_yy+k_zz)} U_d(\bm r)  \nonumber\\
&=  \frac{4 \pi p^2}{3} \left( \frac{3k_z^2}{k_x^2 + k_y^2 + k_z^2} -  1\right)\,.\label{eq:uft}
\end{align}
Comparing Eq. \eqref{2D:2} and Eq. \eqref{2D:3} leads to identity
and
\begin{align}
\tilde V_{2D}({\bm k}) = \int \frac{d k_z}{2 \pi} \, \tilde A(-k_z)  \tilde U(k_x, k_y, k_z)\,,
\label{analytics}
\end{align}
which can be analytically evaluated. It results that $\tilde V_{2D}({\bm k})=\tilde V_{2D}(|{\bm k}|)$, and in detail
\begin{align}
\tilde V_{2D} (q) &=  \int_{q=\sqrt{k_x^2 + k_y^2}}\frac{d k_z}{2 \pi}\, \tilde A(k_z) \tilde U(k_x, k_y, k_z) \\
&=\frac{ 2 \pi p^2}{\sigma_z} \left[ \frac{2 }{3} \sqrt{\frac{2}{\pi}} - q \sigma_z \, \text{erfcx}(q \sigma_z / \sqrt{2})  \right]\,,
\label{correct}
\end{align}
where we used $q=\sqrt{k_x^2 + k_y^2}$ and  $\text{erfcx}(x)$ is the scaled complementary error function: $\text{erfcx}(x) = e^{x^2} \text{erfc}(x)$ \cite{Abramowitz}.  The
 expression in Eq. \eqref{correct} is identical to the one in Ref. \cite{Babadi}, except for the constant term, which modifies the on-site interaction \cite{Erratum}. In real-space it reads
\begin{widetext}
\begin{eqnarray}
V_{2D}(x,y) = p^2 \left[ \frac{e^{\frac{\rho^2}{4 \sigma_z^2}}}{\sqrt{8 \pi} \sigma_z^5} \left( (\rho^2 + 2 \sigma_z^2) \text{K}_0\left(\frac{\rho^2}{4\sigma_z^2}\right) - \rho^2 \text{K}_1\left(\frac{\rho^2}{4 \sigma_z^2}\right)\right) - \frac{\sqrt{2 \pi}}{3 \sigma_z} \delta(x,y)\right].
\label{eq:Recati}
\end{eqnarray}
where $K_{0,1}(x)$ are modified Bessel function of second kind \cite{Abramowitz} and $\rho=|{\bm \rho}|$. We note that the last term in Eq. \eqref{eq:Recati} is an effective attractive contact interaction that can substantially modify the onsite coefficient of the multi-mode EBH model. 
\end{widetext}


\begin{thebibliography}{99}

\bibitem{Lahaye}
T Lahaye, C Menotti, L Santos, M Lewenstein, and T Pfau, Rep. Prog. Phys. {\bf 72}, 126401 (2009).

\bibitem{Jin}
S. A. Moses, J. P. Covey, M. T. Miecnikowski, D.  S. Jin, S. , and J. Ye, Nature Physics, {\bf 13}, 13 (2017).

\bibitem{Ferlaino2016}
S. Baier, M. J. Mark, D. Petter, K. Aikawa, L. Chomaz, Z. Cai, M. Baranov, P. Zoller, and F. Ferlaino,
Science {\bf 352}, 201 (2016).

\bibitem{BlochRMP}
I. Bloch, J. Dalibard, and W. D. Zwerger, Rev. Mod. Phys. {\bf 80}, 885 (2008).

\bibitem{EBH}
T. D. Kuhner, S. R. White, and H. Monien, Phys. Rev. B {\bf 61}, 12 474 (2000).

\bibitem{Bose-Hubbard}
D. Jaksch, C. Bruder, J. I. Cirac, C. W. Gardiner, and P. Zoller, Phys. Rev. Lett. {\bf 81}, 3108 (1998).

\bibitem{FisherAndFisher}
M. P. A. Fisher, P. B. Weichman, G. Grinstein, and D. S. Fisher, Phys. Rev. B {\bf 40}, 546 (1989).

\bibitem{Goral}
K. G\'oral, L. Santos, and M. Lewenstein, Phys. Rev. Lett. {\bf 88}, 170406 (2002).

\bibitem{Menotti}
C. Menotti, C. Trefzger, and M. Lewenstein, Phys. Rev. Lett. {\bf 98}, 235301 (2007).

\bibitem{Batrouni2013}
G. G. Batrouni, R. T. Scalettar, V. G. Rousseau, and B. Gr\'emaud, Phys. Rev. Lett. {\bf 110}, 265303 (2013).

\bibitem{Deng2013a}
Xiaolong Deng, R. Citro, E. Orignac, A. Minguzzi, and L. Santos, Phys. Rev. B {\bf 87}, 195101 (2013).

\bibitem{Deng2013b}
Xiaolong Deng, R. Citro, E. Orignac, A. Minguzzi, and L. Santos, New J. Phys. {\bf 15}, 045023 (2013).

\bibitem{Batrouni2014}
G. G. Batrouni, V. G. Rousseau, R. T. Scalettar, and B. Gr\'emaud, Phys. Rev. B {\bf 90}, 205123 (2014).

\bibitem{Astrakharchik:2007}
G. E. Astrakharchik, J. Boronat, I. L. Kurbakov, and Yu. E. Lozovik,
Phys. Rev. Lett. {\bf 98}, 060405 (2007).

\bibitem{Buechler:2007}
H. P. B\"uchler, E. Demler, M. Lukin, A. Micheli, N. Prokof'ev, G. Pupillo, and P. Zoller,
Phys. Rev. Lett. {\bf 98}, 060404 (2007).

\bibitem{Astrakharchik}
G. E. Astrakharchik, G. Morigi, G. De Chiara, and J Boronat,
Phys. Rev. A {\bf 78}, 063622 (2008).

\bibitem{Altman}
%Nonlocal order in elongated dipolar gases
J. Ruhman, E. G. Dalla Torre, S. D. Huber, and E. Altman,
Phys. Rev. B {\bf 85}, 125121 (2012).

\bibitem{Fishman}
S. Fishman, G. De Chiara, T. Calarco, and G. Morigi,
Phys. Rev. B {\bf 77}, 064111 (2008).

\bibitem{Shimshoni}
E. Shimshoni, G. Morigi, and S. Fishman,
Phys. Rev. Lett. {\bf 106}, 010401 (2011); Phys. Rev. A {\bf 83}, 032308 (2011).

\bibitem{Cartarius}
F. Cartarius, G. Morigi, and A. Minguzzi,
Phys. Rev. A {\bf 90}, 053601 (2014).

\bibitem{Sinha}
S. Sinha and L. Santos,
Phys. Rev. Lett. {\bf 99}, 140406 (2007). 

\bibitem{Deuretzbacher}
F. Deuretzbacher, J. C. Cremon, and S. M. Reimann,
Phys. Rev. A {\bf 81}, 063616 (2010).

\bibitem{Sowinski}
%Dipolar Molecules in Optical Lattices
T. Sowi\'nski, O. Dutta, P. Hauke, L. Tagliacozzo, and M. Lewenstein,
Phys. Rev. Lett. {\bf 108}, 115301 (2012).

\bibitem{Recati}
N. Bartolo, D. J. Papoular, L. Barbiero, C. Menotti, and A. Recati,
Phys. Rev. A {\bf 88}, 023603 (2013). 

\bibitem{Silvi}
P. Silvi, G. De Chiara, T. Calarco, G. Morigi, and S. Montangero,
Annalen der Physik {\bf 525}, 827 (2013);
%Characterization of the quantum linear-zigzag transition using DMRG
P. Silvi, T. Calarco, G. Morigi, and S. Montangero
Phys. Rev. B {\bf 89}, 094103 (2014).

\bibitem{Podolsky}
D. Podolsky, E. Shimshoni, P. Silvi, S. Montangero, T. Calarco, G. Morigi, and S. Fishman,
Phys. Rev. B {\bf 89}, 214408 (2014). 

\bibitem{Abramowitz}
M. Abramowitz and I. Stegun, {\it Handbook of Mathematical
Functions}, (Dover Publications Inc., New York, 1968).

\bibitem{Berg}
E. G. Dalla Torre, E. Berg, and E. Altman, Phys. Rev. Lett. {\bf 97}, 260401 (2006).

\bibitem{Naegerl}
T. Takekoshi, L. Reichs\"ollner, A. Schindewolf, J. M. Hutson, C. Ruth Le Sueur, O. Dulieu, F. Ferlaino, R. Grimm, and H.-C. N\"agerl,
Phys. Rev. Lett. {\bf 113}, 205301 (2014).

\bibitem{Fidelity}
Wen-Long You, Ying-Wai Li, and Shi-Jian Gu,
Phys. Rev. E {\bf 76}, 022101 (2007); M. Cozzini, R. Ionicioiu, and P. Zanardi,
Phys. Rev. B {\bf 76}, 104420 (2007); P. Buonsante and A. Vezzani,
Phys. Rev. Lett. {\bf 98}, 110601 (2007).

\bibitem{Kollath}
C. Kollath, J. S. Meyer, and T. Giamarchi,
Phys. Rev. Lett. {\bf 100}, 130403 (2008).

\bibitem{Knap}
M. Knap, E. Berg, M. Ganahl, and E. Demler,
Phys. Rev. B {\bf 86}, 064501 (2012).

\bibitem{Wall2013}
M. L. Wall and L. D. Carr, New J. of Phys. {\bf 15}, 123005 (2013).
 
\bibitem{Babadi}
M. Babadi and E. Demler, Phys. Rev. A {\bf 86}, 063638 (2012)

\bibitem{Erratum}
M. Babadi and E. Demler, Phys. Rev. A {\bf 87}, 039903 (2013).
\end{thebibliography}
\end{document}